\documentclass[numberedappendix]{emulateapj}
\usepackage{apjfonts}
\newcommand{\rg}{R_{\rm g}}
\shorttitle{GRB060218 as a tidal disruption}
\shortauthors{Shcherbakov, Pe'er, Reynolds, Haas, Bode, \& Laguna}

\begin{document}
\title{GRB060218 AS A TIDAL DISRUPTION OF A WHITE DWARF BY AN INTERMEDIATE MASS BLACK HOLE}

\author{Roman V. Shcherbakov\altaffilmark{1,2,3}, Asaf Pe'er\altaffilmark{4,5}, Christopher S. Reynolds\altaffilmark{1,2},
Roland Haas\altaffilmark{6,7}, Tanja Bode\altaffilmark{7}, Pablo Laguna\altaffilmark{7}}

\altaffiltext{1}{Department of Astronomy, University of Maryland, College Park, MD 20742, USA}
\altaffiltext{2}{Joint Space Science Institute, University of Maryland, College Park MD 20742, USA}
\altaffiltext{3}{Hubble Fellow}
\altaffiltext{4}{Harvard-Smithsonian Center for Astrophysics, 60 Garden Street, Cambridge, MA 02138, USA}
\altaffiltext{5}{Department of Physics, University College Cork, Cork, Ireland}
\altaffiltext{6}{Theoretical AstroPhysics Including Relativity, California Institute of Technology, Pasadena, CA 91125, USA}
\altaffiltext{7}{Center for Relativistic Astrophysics, School of Physics, Georgia Institute of Technology, Atlanta, GA 30332, USA}

\begin{abstract}
A highly unusual pair of a gamma-ray burst (GRB) GRB060218 and an associated supernova SN2006aj has puzzled theorists for years.
A supernova shock breakout and a jet from a newborn stellar mass compact object were put forward to explain its multiwavelength signature.
We propose that the source is naturally explained by another channel, a tidal disruption of a white dwarf (WD) by an intermediate mass black hole (IMBH).
The tidal disruption is accompanied by a tidal pinching, which leads to the ignition of a WD and a supernova.
Some debris falls back onto the IMBH, forms a disk, which quickly amplifies the magnetic field, and launches a jet.
We successfully fit soft X-ray spectrum with the Comptonized blackbody emission from a jet photosphere.
The optical/UV emission is consistent with self-absorbed synchrotron from the expanding jet front.
The accretion rate temporal dependence $\dot{M}(t)$ in a tidal disruption provides a good fit to soft X-ray lightcurve.
The IMBH mass is found to be about $10^4M_\odot$ in three independent estimates: (1) fitting tidal disruption $\dot{M}(t)$ to soft X-ray lightcurve;
(2) computing the jet base radius in a jet photospheric emission model; (3) inferring the central BH mass based on a host dwarf galaxy stellar mass.
The supernova position is consistent with the center of the host galaxy, while low supernova ejecta mass is consistent with a WD mass.
High expected rate of tidal disruptions in dwarf galaxies is consistent with one source observed by \textit{Swift} satellite
over several years at GRB060218 distance of $150$~Mpc. The encounters with the WDs provide a lot of fuel for IMBH growth.
\end{abstract}


\section{INTRODUCTION}
The existence of the stellar mass black holes (BHs) with mass $M_{BH}<100M_\odot$ and the supermassive black holes (SMBH)
with mass $M_{BH}>10^5M_\odot$ has long been established.
A population of intermediate mass black holes (IMBH) likely exists. Those BHs have masses in a range $100M_\odot<M_{BH}<10^5M_\odot$.
They could live in the centers of dwarf galaxies \citep{Dong:2007dg,Greene:2012kr} or globular clusters (GCs)
\citep{Fabbiano:1997yt,Fabbiano:2001fr,Colbert:1999fe,Matsumoto:2001oi,Gultekin:2004gh}.
The IMBHs can form by a collapse of a massive cloud \citep{Begelman:2006jk} or a massive star \citep{Fryer:2001fe,Madau:2001po,Schneider:2002dp} or grow from a stellar mass BH.
A star cluster G1 in M31 galaxy is estimated to host a $2\times10^4M_\odot$ BH \citep{Gebhardt:2002hj} based on the velocity dispersion profile.
SDSS J160531.84+174826.1 dwarf galaxy is estimated to have an IMBH with a mass $\sim7\times10^4M_\odot$ based on luminosity scaling relations \citep{Dong:2007dg}.
One of the best candidates is HLX-1 source in ESO 243-49 galaxy. Based on a thin disk thermal state the estimated BH mass is $3\times10^4M_\odot$ \citep{Davis:2011ka}.
All those candidates are still tentative and the uncertainty of the mass estimates is up to an order of magnitude.
More IMBH candidates with qualitatively different observational signatures may provide stronger evidence for existence of such objects.

Tidal disruptions of the white dwarfs (WDs) by IMBHs provide such qualitatively different signature.
Unlike disruptions of main sequence (MS) stars, which happen far away from the BHs and have slow timescales, the disruptions of WDs
are very fast. Such disruptions may lead to accretion rates up to $10^{4}M_\odot{\rm yr}^{-1}$ \citep{Haas:2012ak}.
A flow with an extreme accretion rate may produce a short powerful burst of radiation, for example, when a jet is launched.
The X-ray emission of \textit{Swift} J1644+57 source is attributed to a jet launched by a super-Eddington accretion disk formed after a tidal disruption
of a star by a SMBH \citep{Bloom11,Burrows11,Levan2011,Zauderer11}. An alternative theory \citep{Krolik:2011ax} ascribes \textit{Swift} J1644+57
to an encounter of a WD with an IMBH.

When particle acceleration is inefficient, then the only radiation from the jet is the photospheric emission \citep{Eichler:2000op,Meszaros:2000le,Daigne:2002oa,Rees:2005lk}.
The radiation field is in a thermal equilibrium with matter within a dense hot jet up to the distance $R_{\rm ph}$ along the jet, called a photospheric radius, where the optical depth
to Compton scattering is about unity $\tau_\sigma\sim1$. The photospheric emission has a quasi-blackbody spectrum \citep{Peer:2007sd,Peer:2008pa,Peer:2011jq,Beloborodov:2011pl}.
Particle acceleration may happen either in the entire volume of the jet or in the internal shocks.
Dissipation in the volume of the jet, in particular, in the sub-photospheric regions leads to a modified blackbody spectrum from near the jet photosphere
\citep{Peer:2005tu,Peer:2006ph,Giannios:2006je,Giannios:2012ad}.
Modifications include the Comptonization and the broadening of the blackbody peak with pair production.
Effective particle acceleration in the internal shocks \citep{Meszaros2006} is likely responsible for the emission of gamma-ray bursts (GRBs).
When two shells moving at different speeds collide above the photosphere, a shock forms, and a substantial fraction of relative kinetic energy can be transferred into the electrons
\citep{Rees:1994fa,Sari:1997ge}, which radiate synchrotron emission. The collisions of shells are especially efficient, when the jet bulk Lorentz factor $\Gamma$ is large.
For the small Lorentz factors and the large photospheric radii, such as in the jets from tidal disruptions of the WDs by the IMBHs, the collisions of shells may happen within the photosphere,
and the internal shock signatures might be weaker. The fluctuations of the jet $\Gamma$ diffuse out, while the jet travels within the photosphere, and the fluctuations
outside of the photosphere are small. Even when the slow shells collide, the energy release is weak.
Therefore, we expect to see the strong blackbody signature of the photospheric emission. While the energy density of the slow tidal disruption jet might not warrant fast pair production,
the Comptonization is still expected to modify the spectrum.

A distinct feature of the WD disruptions by IMBHs is a supernova. As a consequence of the tidal compression along the angular momentum axis, the WD may undergo thermonuclear ignition
\citep{Luminet:1985re,Luminet:1985yt,Luminet:1989th} and explode.  The energy and the composition of the resulting supernova are the functions
of the WD mass $m_{WD}$, the pericenter radius $R_P$, and the BH mass $M_{BH}$ \citep{Rosswog:2009ie}.
If the WD is massive and the disruption is deep, then the explosion could be similar to a supernova Type Ia with a comparable energy release.
Less massive WDs may lead to low luminosity explosions with little $Ni$ synthesized. There is no explosion at all in some cases.
In any case the ejected mass $\lesssim1M_\odot$ is less than in the core-collapse supernovae.

In sum, we predict a transient similar to a GRB, but softer and longer, accompanied by a supernova with a small ejecta mass.
The best candidate we find in \textit{Swift} GRB catalogue is GRB060218 source.
This unusual event is an underluminous very long GRB with a duration $t_{90}\approx2600$~s and with a smooth X-ray lightcurve \citep{Soderberg:2006na}.
It is accompanied by a fast supernova SN2006aj, which was modeled to have a low ejecta mass $M_{\rm ej}\sim(1-2)M_\odot$ \citep{Mazzali:2006na}.
The X-ray emission has a blackbody component characteristic of the photospheric radiation.
The early X-ray radiation is accompanied by the powerful optical/UV emission \citep{Ghisellini:2007bb}.
Two classes of theories were proposed to explain the source: a supernova shock breakout model \citep{Campana2006,Waxman:2007ap,Nakar:2012aj}
and a model with a mildly relativistic jet from a newborn compact object, such as a magnetar \citep{Fan:2006ag,Toma2007,Ghisellini:2007aa,Ghisellini:2007bb}.
Nevertheless, we think that a tidal disruption of a WD by an IMBH is not only a viable model for the source, but also more naturally explains some features,
such as the duration and the soft quasi-thermal spectrum.

The paper is organized as follows. In \S~\ref{sec:grbobs} we describe in more detail the observations of GRB060218/SN2006aj. We extensively discuss
in \S~\ref{sec:former} the former theoretical explanations of the source: the shock breakout model and the jet launched by the newborn compact object.
Then in \S~\ref{sec:tidal} we concentrate on modeling within a tidal disruption scenario. We rederive the dynamics of a tidal disruption and
discuss the physics of jet launching. We perform time-resolved spectroscopy of the source and successfully model the X-ray emission
with a Comptonized blackbody spectrum from the photosphere. The derived jet base radius corresponds to about $10^4M_\odot$ IMBH.
We propose that the origin of the powerful early optical/UV radiation is the front region of propagating jet.
We find that the full X-ray lightcurve can be fitted well by a scaled dependence of the fallback accretion rate on time $\dot{M}(t)$.
The fit provides the IMBH mass about $10^4M_\odot$. We find that the steep decay phase is consistent with the action of absorption alone and that the afterglow
can be naturally explained as powered by the central engine operating in a shallow $t^{-4/3}$ regime. The associated supernova is consistent with a WD origin,
while its position is consistent with the center of a host galaxy. The mass of the dwarf host galaxy provides the estimate of a central IMBH mass about $10^4M_\odot$.
In \S~\ref{sec:rates} we estimate the event rates. A high rate of tidal disruptions is predicted in the dwarf galaxies,
while the disruptions of the WDs constitute a significant percentage of all disruptions. Since the disruptions of the MS stars mostly happen at large distances from the BHs,
then such events could be much dimmer, than the disruptions of the WDs.
The tidal disruptions of the WDs provide a plenty of material to feed the central IMBH up to a supermassive size.

\section{OBSERVATIONS OF GRB060218}\label{sec:grbobs}
GRB060218 triggered the Burst Alert Telescope (BAT) onboard \textit{Swift} mission satellite on 18 Feb 2006 \citep{Campana2006}.
Soon after the trigger the X-ray Telescope (XRT) identified a bright source, whose count rate peaked
at around $960$~s from BAT initial trigger.
Then the source gradually decayed over a continuous observation period, which ended at around $2700$~s.
The event duration $t_{90}=2600$~s is unusually long among \textit{Swift} GRBs \citep{Campana2006}. The XRT lightcurve was unusually smooth and regular.
The rise period is characterized by hard emission with the BAT flux about equal to the XRT flux \citep{Campana2006,Toma2007} and with the peak energy $E_p\sim5$~keV.
The isotropic-equivalent luminosity is $L_{\rm iso}\sim10^{47}{\rm erg~s}^{-1}$, which is about $10^5$ times less than for the typical GRBs \citep{Toma2007}.
The source is found to have a blackbody component with a low temperature $T\sim0.2$~keV, whose flux contribution increased with time from $15\%$ till $80\%$ \citep{Campana2006,Li:2007sh}.
The \textit{Swift} satellite returned to observe the source at $\sim6000$~s and found it in a steep decay phase with the absorbed flux $\sim100$ times below the peak value.
After about $1.5\times10^4$~s the decay of the X-ray flux flattened into a shallow afterglow with the X-ray luminosity $L_X\propto t^{-1.2}$ \citep{Soderberg:2006na}.
The simultaneous observations by the Ultra-Violet/Optical Telescope (UVOT) onboard \textit{Swift} satellite revealed substantial early emission peaking at about $5\times10^4$~s with
a dereddened UV flux $\nu F_\nu\sim10^{-10}{\rm erg~s}^{-1}{\rm cm}^{-2}$ \citep{Ghisellini:2007bb}. A lower peak at a level $\nu F_\nu\approx6\times10^{-12}{\rm erg~s}^{-1}{\rm cm}^{-2}$
followed after several days. The second peak is attributed to a supernova SN2006aj \citep{Maeda:2007gh,Mazzali:2006na,Modjaz:2006ua,Pian:2006na}, while the origin of the first peak
is debated \citep{Waxman:2007ap,Ghisellini:2007aa,Ghisellini:2007bb}.

The supernova was classified as Type Ic \citep{Mazzali:2006na}. However, its unique spectral properties prompted
to suggest a new Type Id classification \citep{Mazzali:2006na,Maeda:2007gh}. The rapid supernova peaked at $10$~days \citep{Pian:2006na},
which is the fastest of all supernovae associated with GRBs \citep{Ferrero:2006}.
According to \citet{Mazzali:2007as}, "the ejected mass predicted below $8000{\rm km~s}^{-1}$  by the model
used by \citet{Mazzali:2006na} is $M_{\rm ej}=1M_\odot$", while the nebular phase observations suggest a higher total ejecta mass
$M_{\rm ej}=2M_\odot$ \citep{Mazzali:2006na,Mazzali:2007as,Maeda:2007gh}. The estimated ${}^{56}$Ni mass in the ejecta is $0.2M_\odot$.
The ejecta consists mostly of oxygen and carbon \citep{Mazzali:2007as}, but also contains some silicon and iron \citep{Mirabal:2006oa}.
The correspondent supernova energy is $E_K\sim2\times10^{51}{\rm erg}$.
Both the ejecta mass and and total energy are much less than the typical values $M_{\rm ej}=10M_\odot$  and $E_K=3\times10^{52}{\rm erg}$
for GRB-supernovae (see \citealt{Mazzali:2006na} and references therein). Substantial optical polarization \citep{Gorosabel:2006hj}
indicates the asymmetry of the ejected material.

The optical observations of the host galaxy indicate a source redshift $z=0.0335$ \citep{Mirabal:2006oa,Modjaz:2006ua}, which corresponds to a distance $d=143$~Mpc.
GRB060218 is associated with a dwarf star-forming galaxy with the stellar mass $M_{\rm st}\sim10^{7.2}M_\odot$ \citep{Ferrero:2007jd},
the metallicity $Z\approx0.07Z_\odot$ \citep{Wiersema2007}, and the characteristic radius $R_{80}=0.55$~kpc \citep{Svensson:2010ka}.
The supernova and the host galaxy were observed by \textit{Hubble} Advanced Camera for Surveys (ACS) instrument under the program GO 10551 (PI Kulkarni) in cycle 14.
The images produced by \citet{Misra:2011jg} reveal some irregularity of the host dwarf galaxy morphology.

\section{FORMER THEORETICAL MODELS}\label{sec:former}
GRB060218 and the associated supernova SN2006aj sparked substantial interest among the researchers with over ten papers being dedicated to the theoretical explanations
of this highly unusual source. The theoretical efforts can be divided into two big categories: a shock breakout model and a model with a jet
launched by a magnetar or by a stellar mass BH. In the following subsections we briefly review the theoretical models emphasizing their successes and problems.
In the following section we offer an explanation of the source within the tidal disruption scenario.
\subsection{Shock Breakout Model}
The shock breakout model states that the nuclear explosion following the collapse of a massive core launches a shock wave, which propagates out through the star \citep{Colgate:1968dw}.
The radiation-dominated shock deposits its energy into the low-density gas at the stellar surface \citep{Katz2010}.
The heated gas radiates bremsstrahlung photons, which are inverse-Compton scattered into the X-ray and the $\gamma$-ray bands. Gas exponentially cools down.
The shock deposits up to $10^{48}{\rm erg}$ of energy, most of which is radiated as an X-ray flash \citep{Colgate:1974ja}.
A variety of sources were successfully modeled with the supernova shock breakout.
A good example is XRF080109 and the associated Type Ib/c supernova SN2008D. The large supernova ejecta mass for that source $M_{\rm ej}=4-7M_\odot$ \citep{Soderberg:2008dq,Mazzali:2008dw}
and the low X-ray radiation energy $10^{45}-10^{46}{\rm erg}$ \citep{Chevalier:2008fe} leave little doubts about its shock breakout origin.
Other sources include the X-ray brightening of SNLS-04D2dc supernova \citep{Schawinski:2008dw} with low X-ray radiated energy $\le10^{47}{\rm erg}$,
SN1998bw with the ejecta mass $\sim12M_\odot$ accompanied by XRF/GRB980425 \citep{Woosley:1999la},
SN2003lw with the ejecta mass $13M_\odot$ accompanied by a weak XRF/GRB031203 \citep{Mazzali:2006pa}.

In the case of XRF060218 the deposited energy $10^{49.5}{\rm erg}$ is quite large \citep{Campana2006}.
The peak isotropic unabsorbed flux of the soft X-ray component is $F_{BB}=(3-6)\times10^{-8}{\rm erg~s}^{-1}{\rm cm}^{-2}$ with the temperature $T_{BB}=0.11-0.17$~keV \citep{Butler:2007uy}.
These parameters correspond to blackbody emission from the non-relativistic medium at a radius  $R=(3-8)\times10^{12}$~cm,
while \citet{Campana2006} inferred $R=(0.5-1)\times10^{12}$~cm within their analysis at the early times and $R=(0.2-2)\times10^{12}$~cm at the late times.
This radius indicates a compact progenitor such as a Wolf-Rayet (WR) star.
Large explosion energy might be inconsistent with the non-relativistic expansion.
Mildly relativistic ejecta velocity with $v\approx0.85c$ was proposed in later modeling by \citet{Waxman:2007ap},
who also stated that the emission should come from the photosphere above the stellar surface.
The relativistic shock breakout theory was further developed by \citet{Nakar:2012aj}, who computed the shock dynamics, the lightcurve and the spectrum.
The spectrum is predicted to be quasi-thermal with the temperature around $50$~keV.
\citet{Nakar:2012aj} derived a relation between the XRF event duration, the total energy, and the observed temperature
\begin{equation}\label{eq:breakout}
t_{bo}\sim 20{\rm s}\left(\frac{E_{bo}}{10^{46}{\rm erg}}\right)^{1/2}\left(\frac{T_{bo}}{50{\rm keV}}\right)^{-2.68}.
\end{equation}
To satisfy this relation the early temperature of $40$~keV is taken for GRB060218 consistent with the peak energy $36$~keV at the very early times \citep{Toma2007}.

The shock breakout theory has its problems. \citet{Li:2007sh} concluded that the observed temperature and the total energy of GRB060218 lead
to the unrealistically large photospheric radius $R_{\rm ph}$ of a WR star inconsistent with the galactic WR stellar population.
This conclusion is strengthened by our larger estimate of the emission radius.
The large $R_{\rm ph}$ in the model by \citet{Nakar:2012aj} corresponds to unrealistically dense stellar wind.
\citet{Ghisellini:2007aa} showed that the optical/UV spectrum of the source is too bright for the X-rays and the optical to be the parts of the same blackbody component.
In response, \citet{Waxman:2007ap} "fiercely argued"\footnote{\citep{Ghisellini:2007bb}} that the shock breakout is anisotropic,
and the different regions emit at the different wavelengths, which helps to reconcile the theory with the observations.
The prompt thermal X-rays are emitted by a compressed shell, while the optical radiation originates in the outer shells of the expanding star at a much larger radius.
In their final reply, \citet{Ghisellini:2007bb} found the anisotropic expansion unconvincing. The simple energetics argument is dramatically inconsistent with
the shock breakout model by \citet{Waxman:2007ap}. The isotropic optical thermal emission at $t\sim10^3$~s
requires the temperature about $\sim1\times10^6$~K, which corresponds to the total energy carried by a shock above $10^{51}$~erg, much in excess of any shock breakout model.

\subsection{Jet Launched by a Magnetar or a Stellar Mass BH}
A promising alternative explanation for GRB060218 is the jet launched by the central engine.
A newborn central engine, a magnetar or a BH, results from a core collapse. The jet pierces through the star and escapes.
That is how a typical GRB operates and our source can be just on a low-luminosity end of the spectrum.
A jet with a low Lorentz factor $\Gamma\sim5$ and a wide opening angle $\theta\sim0.3$ can produce the emission,
which peaks in the X-rays and exhibits no jet break \citep{Toma2007}. Such a jet may be powered over a long timescale by a magnetar.
The central engine activity is the natural explanation of the afterglow \citep{Soderberg:2006na,Fan:2006ag}.
\citet{Dainotti:2007} proposed a model of the interaction of a electron-positron fireshell and circumburst medium to explain the afterglow.
Their best model constrains the density profile to be $n\propto r^{-\beta}$, where $\beta=1.0-1.7$ up to $10^{18}$~cm.
The unusually bright prompt optical/UV emission can be explained as produced by the self-absorbed Comptonized synchrotron \citep{Ghisellini:2007aa},
while the blackbody  X-rays result from the photospheric jet emission \citep{Ghisellini:2007bb}.
\citet{Bufano:2012dd} argued that the central magnetar would spin down too rapidly and could not power the source over thousands of seconds,
while expending most of its energy in the expansion of the dense envelope. Yet, \citet{Quataert11} proposed a model in application to \textit{Swift} J1644+57,
where a magnetar powers a GRB source over many days or even weeks.

The relativistic jet is a natural explanation for the source. However, if GRB060218 was a GRB, then it clearly was an unusual one.
The long GRBs have large total energies $E_{\rm tot,iso}=10^{52-54}$~erg \citep{Nava:2008kp}, short durations $t_{90}=2-200$~s
\citep{Butler:2007op}, and higher Lorentz factors $\Gamma=100-1000$ \citep{Peer:2007sd,Liang:2010pa,Ghisellini:2012grb}.
The high Lorentz factor of a typical long GRB is the consequence of jet confinement by the pressure from the star \citep{Tchekhovskoy:2010na}.
\citet{Bromberg:2011aw} argued that the GRB population does not extend to the slow low-luminosity end.
Thus, another origin of events such as GRB060218 is to be sought. Below we propose that the jet is launched by a tidal disruption.
This model provides the explanations for the observed event duration, the spectrum, and the lightcurve.
\section{TIDAL DISRUPTION MODEL}\label{sec:tidal}
GRB060218 can be best modeled by a low-luminosity wide mildly relativistic outflow/jet. The tidal disruption of a WD by an IMBH produces such a jet.
In this section we discuss the expected temporal and spectral properties of such tidal disruptions,
fit observations of GRB060218, and consider the accompanying supernova SN2006aj and the host galaxy.
Extensive observations of the source pose multiple tests for any theory aimed to explain it.
The tidal disruption scenario passes all these tests.

\subsection{Disruption Dynamics}\label{subsec:dynamics}
\subsubsection{Fallback Material and Disk Formation}
A star gets tidally disrupted, when it approaches sufficiently close to the BH.
Let us define a tidal radius as
\begin{equation}\label{eq:Rtidal}
R_T=\left(\frac{2M_{BH}}{m_{WD}}\right)^{1/3}R_\star.
\end{equation}
Only the stars on orbits with the pericenter distance $R_P$ smaller than about
\begin{equation}\label{eq:RP}
R_P\lesssim R_T
\end{equation} can get disrupted (see \citealt{Evans:1989qe} and references therein).
The WD radius is \citep{Naunberg:1972dw}
\begin{equation}\label{eq:Rstar}
R_\star=8.5\times10^8{\rm cm}\left(\frac{m_{WD}}{M_{\rm Ch}}\right)^{-1/3}\left[1-\left(\frac{m_{WD}}{M_{\rm Ch}}\right)^{4/3}\right]^{1/2},
\end{equation} where $M_{\rm Ch}=1.44M_\odot$ is the Chandrasekhar mass. We calibrated the normalization to reproduce a typical observed WD radius \citep{Nalezyty:2004ca}
for one solar mass $m_{WD}=1M_\odot$. A tidal radius depends weakly on the BH mass, while the event horizon radius $\rg=G M_{BH}/c^2$ is proportional to the BH mass.
Thus, the SMBHs will swallow the WDs without disruptions and only the BHs with relatively small masses can disrupt the WDs \citep{Luminet:1989yo,Rosswog:2009ie}.
The stars are swallowed, if
\begin{equation}\label{eq:Rswallow}
R_P<f \rg,
\end{equation} where $f\approx8$ for a non-spinning BH, which captures massive geodesics with specific angular momentum $\tilde{l}<4\rg c$ \citep{Shapiro1986}.
The value of $f$ can be much lower for spinning BHs \citep{Kesden:2012a}. Combining equations (\ref{eq:Rtidal},\ref{eq:RP},\ref{eq:Rswallow}) we find that only the BHs with mass
\begin{equation}
M_{BH}\lesssim\frac{\sqrt{2}c^3}{\sqrt{M_\star}}\left(\frac{R_\star}{f G}\right)^{3/2}
\end{equation} can tidally disrupt a star. The correspondent critical mass of a non-spinning BH to disrupt $0.8M_\odot$ WD is $3\times10^4M_\odot$.
For a star on a parabolic orbit about $50\%$ of its material ends up being bound to the BH and about $50\%$ remains unbound in a standard picture \citep{Evans:1989qe}.
The fractions may change, when the pericenter distance is close to the BH marginally bound orbit radius (several gravitational radii).
Among models with different pericenter radii, BH spins, and orientations \citet{Haas:2012ak} found the cases, when almost all material is captured or almost all material remains unbound.

The bound material makes its way onto the BH. In a classic theory the fallback time is \citep{Evans:1989qe}
\begin{equation}
t_{\rm fb}\sim \frac{R_P^3}{\sqrt{G M_{BH}}R_\star^{3/2}}
\end{equation} and the fallback rate of debris is
\begin{equation}\label{eq:Mdot}
\dot{M}_{fb}=\frac13\frac{M_\star}{t_{fb}}\left(\frac{t_{fb}}{t}\right)^{5/3}.
\end{equation}
The fallback time $t_{\rm fb}$ can be anywhere from several hours down to $3$~min \citep{Haas:2012ak} for the disruptions of the WDs by the IMBHs.
The peak accretion rate is achieved at
\begin{equation}\label{eq:tpeak}
t_{\rm peak}=\delta\frac{R_P^3}{\sqrt{G M_{BH}}R_\star^{3/2}},
\end{equation} where $\delta\approx3.33$ for the adiabatic index $\Gamma_{\rm ad}=5/3$ based on simulations by \citet{Evans:1989qe} and \citet{Laguna1993c} for $\beta_T=R_T/R_P=1$,
while $\delta\approx5.5$ for $\Gamma_{\rm ad}=1.4$ \citep{Lodato:2009fr}. The smaller values of the adiabatic index $\Gamma_{\rm ad}<5/3$ are more appropriate for the heavy WDs.
The correspondent peak fallback rate is
\begin{equation}\label{eq:Mdotpeak}
\dot{M}_{\rm peak}\approx0.05m_{WD}\frac{\sqrt{G M_{BH}}R_\star^{3/2}}{R_P^3}
\end{equation} for $\Gamma_{\rm ad}=1.4$.
The fallback rate rises from zero, when the most bound debris just reach the BH, up to the maximum value $\dot{M}_{\rm peak}$
at $t_{\rm peak}$ and then decreases according to $t^{-5/3}$ law.

The fallback matter forms an accretion disk at a circularization radius $R_{\rm circ}=\eta R_P$, where $\eta=2$ according to the conservation of angular momentum.
The disk exists as a radiatively inefficient accretion flow (RIAF), since the material density is very high and the photons cannot escape \citep{Abramowicz1988}.
The fallback rate determines the accretion rate during the early evolution of the system.
After the disk is formed, it starts to evolve on a slow viscous timescale for $R_{\rm circ}$ radius \citep{Cannizzo1990}.
Such evolution changes the temporal slope of the accretion rate to $t^{-4/3}$ for adiabatic RIAFs \citep{Cannizzo11}.
RIAFs with the outward energy flux may unbind the material in the outer disk \citep{Narayan:2001fe,Metzger:2012ww}.
Then only a small percentage of the material reaches the BH. However, the inner disk regions
with the radius less than $\sim20\rg$ evolve fast and are not influenced by relatively slow energy transport \citep{Abramowicz:2002hg}.

As was recently shown by general relativistic magneto hydrogynamic (GRMHD) simulations \citep{Tchekhovskoy:2011qp,McKinney2012},
a poloidal magnetic field is necessary to launch a jet. The field strength for a strong jet can be estimated based on the equipartition argument.
When the magnetic field energy density is comparable to the internal energy density of plasma as
\begin{equation}\label{eq:BEdd}
\frac{B_{BH}^2}{8\pi}\sim0.1\frac{\dot{M}c}{4\pi \rg^2}\sim 0.1n_{\rm nuc}m_p c^2,
\end{equation} where $n_{\rm nuc}$ is the density of nucleons, $m_p$ is the proton mass.
Then the magnetic field near the event horizon is
\begin{equation}\label{eq:Beq}
B_{BH}\sim1.3\times10^{10}\left(\frac{\dot{M}}{10^3M_\odot{\rm yr}^{-1}}\right)^{1/2}\left(\frac{M_{BH}}{10^4M_\odot}\right)^{-1}{\rm G}.
\end{equation}
The magnetic field can reach $10^{15}-10^{16}$~G at the base of a GRB jet \citep{Usov:1992er,Wheeler:2000gh,Uzdensky:2007gh,Takiwaki:2009fc}.
A powerful jet is launched after a WD disruption by an IMBH for $B_{BH}\sim10^{10}$~G. How can such a strong poloidal magnetic field be produced?

\subsubsection{Generation of Magnetic Field}
The magnetic field lines get advected towards the BH event horizon with the fallback material.
The magnetic field of a WD is typically within $B\lesssim10^4$~G \citep{Angel:1978yt,Putney:1999dg} with only a few examples of a stronger field.
Since the WD radius is on the order of the BH gravitational radius, then no field amplification is expected due to compression in the converging fallback flow.
The BH may possess an accretion disk with a large magnetic field even before the disruption happens.
When the accretion rate is close to the Eddington rate, the BH sustains the so-called Eddington magnetic field  \citep{Rees:1984de,Daly:2011ra}
\begin{equation}
B_{\rm Edd}=6\times10^6 \left(\frac{M_{BH}}{10^4M_\odot}\right)^{-1/2}{\rm G}.
\end{equation} 
Somewhat larger magnetic field can be accumulated on a BH by infalling debris, which drags pre-existing magnetic flux towards the event horizon.
Such magnetic field generation is a promising mechanism for \textit{Swift} J1644+57 source \citep{Tchekhovskoy:2013ad}.
However, the required $B$-field amplification factor $10^3$ for the tidal disruption of a WD might not be attained by inward dragging of pre-existing magnetic flux.

In situ amplification of the initial seed magnetic field into the equipartition poloidal magnetic field is needed to sustain a powerful jet.
The magneto-rotational instability (MRI) \citep{Balbus:1991jk,Balbus:1998fe} can increase the turbulent magnetic field strength
with an e-folding time of $3$ orbital periods \citep{Stone:1996ed}.
The growth by a $10^3$ factor can be achieved over $\sim25$ local orbits.
The orbital period is \citep{Bardeen1972,Shapiro1986}
\begin{equation}
t_{\rm orb}=\frac{2\pi\rg}c\left[\left(\frac{R_{ISCO}}{\rg}\right)^{3/2}+a^\star\right]\approx 2.5\left(\frac{M_{BH}}{10^4M_\odot}\right){\rm s}.
\end{equation} at the innermost stable circular orbit (ISCO) for the dimensionless spin $a^\star=0.6$.
In sum, a $10^4M_\odot$ BH can amplify the random magnetic field $10^3$ times over as little as $60$~s.

The MRI turbulence produces random $B$-field, while the regular poloidal magnetic field is needed to launch a jet.
The magnetic field generation in a turbulent medium should occur via the dynamo action \citep{Brandenburg:2005pr}, which operates on a viscous timescale
determined by the radial velocity $v_r$ \citep{Davis:2010da,Oneill:2011ad}.
The geometrically thick accretion flow with the effective dimensionless viscosity $\alpha=0.1-0.3$ \citep{King:2007fk} has
the radial velocity $v_r\lesssim\alpha v_K$, where $v_K$ is the Keplerian velocity. Then the viscous timescale at the ISCO can be as short as $15$~s for a $10^4M_\odot$ BH.
Substantial generation of the regular poloidal field component is expected over $\gtrsim10$ viscous timescales or $\gtrsim150$~s.
The failure of the previous attempts to generate the magnetic field via the dynamo action might arise from the low resolution in those simulations
(Jonathan McKinney, private communication). In sum, it might be possible to generate the substantial poloidal magnetic field before the estimated accretion rate peak time.

\subsubsection{Jet Launching}
The jet is a relativistic outflow of the material from near the BH.
The spinning BH surrounded by the accreting magnetized gas is expected to launch a jet \citep{Blandford1977}. We can estimate the jet power via a Blandford-Znajek formula.
When the magnetic field energy density reaches the equipartition with the matter energy density close to the BH, then the Blandford-Znajek jet power
is approximately  \citep{McKinney2005,Tchekhovskoy2010}
\begin{equation}\label{eq:LBZ}
L_{\rm kin}\sim P_0 {a^\star}^2 \dot{M}c^2.
\end{equation} While the simulations with the maximum magnetic flux were able to reach $P_0\sim1$ \citep{Tchekhovskoy:2011qp,McKinney2012}, more traditional values
observed in the simulations with the weak initial field are $P_0=0.01-0.1$ \citep{McKinney:2009rt,Penna:2010dj}.
The tidal disruption of a WD by an IMBH has a typical maximum accretion rate of
$\dot{M}_{\rm peak}\sim10^4M_\odot{\rm yr}^{-1}$ according to the formula~(\ref{eq:Mdotpeak}).
This corresponds to the jet power $L_{\rm kin}\sim 0.01 \dot{M}c^2=5\times10^{48}{\rm erg~s}^{-1}$.
Since there is no theoretical understanding whether or not the dynamo action would quickly generate the poloidal magnetic field,
the kinetic power estimated by the equation~(\ref{eq:LBZ}) is an upper limit. The actual jet power may be substantially lower.

The kinematics of the jet is characterized by the bulk Lorentz factor $\Gamma$.
The GRB jets have very high $\Gamma=100-1000$ likely because of confinement by the surrounding star \citep{Tchekhovskoy:2010na}.
In turn, the active galactic nuclei (AGN) jets are not strongly confined by the ambient gas.
Their $\Gamma$-factors are about $\Gamma\sim10$ \citep{Jorstad:2005aj,Pushkarev:2009ad}.
Similarly, the jets from the low-mass X-ray binaries (LMXBs) have bulk Lorentz factors about $\Gamma\sim10$ \citep{Miller-Jones:2006ma}.
The tidal disruption debris is expected to scatter in radius, cool, and provide little pressure support for the jet regardless of the direction, where the debris is scattered.
Therefore, the tidal disruption jet is expected to have $\Gamma\sim10$.
A low Lorentz factor $\Gamma\sim10$ is indeed suggested for \textit{Swift} J1644+57 tidal disruption event \citep{Metzger11,Liu:2012kw}.

\subsection{Prompt X-ray Emission}\label{subsec:promptX}
Prompt X-ray emission from a slow jet is expected to have a Comptonized blackbody spectrum produced near the jet photosphere.
We perform time-resolved spectroscopy of \textit{Swift} XRT observations of the source and fit the spectrum with the Comptonized blackbody model.
We follow \citet{Butler:2007uy} and cut the full XRT window mode observations into $11$ time slices with about $16,000$ photons in each.
We run \textit{xrtpipeline}, select the time slices with \textit{xselect}, take the appropriate response matrix files (RMF) from the calibration database (CALDB) (version 11)
as indicated by \textit{xrtmkarf} routine within the pipeline, and use the ancillary response files (ARF) generated by \textit{xrtpipeline}.
We run \textit{grppha} to group nearby bins to have at least $20$ photons per bin and perform the minimization of least squares.
We model the spectrum in XSPEC v12.7 \citep{Arnaud:1996oa} with the broken power-law \textit{bknpower}
and with the blackbody spectrum Comptonized by thermal electrons as a part of a \textit{compPS} model \citep{Poutanen:1996jh}.
We favor the exact \textit{compPS} model over the approximate prescriptions for thermal Comptonization of the blackbody spectrum such as \textit{compBB} \citep{Nishimura:1986dw}
and \textit{compTT} \citep{Titarchuk:1994tt}. We compute the model flux in $(0.05-10)$~keV band to capture very soft thermal X-rays.
Unlike \citet{Butler:2007uy} and \citet{Campana2006}, we consider the host galaxy to be a low-metallicity absorber with $Z=0.07Z_\odot$ \citep{Wiersema2007},
while following \citet{Campana2006} we fix the Galactic column at $N_H=9.4\times10^{20}{\rm cm}^{-2}$ with the solar metallicity.
We ignore any potential changes of the hydrogen absorption column with time. We search for a joint best fit to all time slices with a single host galaxy $N_H$.

The results of spectral fitting are presented in Table~\ref{tab:bknpower} for the broken power-law fits. A substantial degeneracy exists between the soft power-law
slope in a \textit{bknpower} model and the host $N_H$. We fix the soft photon index to be $\Gamma_1=-1$, which is representative of the Rayleigh-Jeans tail
$F_\nu\propto\nu^2$ of the blackbody. The host galaxy hydrogen column $N_H=0.750\times10^{22}{\rm cm}^{-2}$ provides the best joint fit
to all time slices with $\chi^2=5463.0$ for ${\rm dof}=4836$ degrees of freedom. The break energy is $0.7-0.8$~keV and depends weakly on the total flux. Since the emission
becomes substantially softer with time, the absorbed flux constitutes a much smaller fraction of the unabsorbed flux at the late times.
This masks the true temporal evolution of the source flux.
\begin{table*}
\scriptsize
\caption{Time-resolved spectroscopy of GRB060218 soft X-ray spectrum. Fitting with a broken power-law.}\label{tab:bknpower}
\begin{centering}
\begin{tabular}{ | p{10mm} | p{20mm}| p{20mm} | p{20mm} | p{32mm} | p{32mm} | p{15mm}|}
\tableline\tableline
  Number  & Time period $t~[{\rm s}]$ & Break energy $E_{\rm break} [{\rm keV}]$ & High energy slope $\Gamma_2$ & Absorbed flux $F_{\rm abs} [10^{-9}{\rm erg~s}^{-1}{\rm cm}^{-2}]$ &
Unabsorbed source flux $F_{\rm unabs} [10^{-9}{\rm erg~s}^{-1}{\rm cm}^{-2}]$ \\
\tableline
1&	164-478&	0.795&	1.226&  4.753&	5.234  \\\tableline
2&	478-691&	0.740& 	1.171&	7.411&	8.148  \\\tableline
3&	691-875&	0.820&	1.234&	8.242&	9.064  \\\tableline
4&	875-1049&	0.746&	1.290&	8.572&	9.575  \\\tableline
5&	1049-1226&	0.813&	1.419&  7.595&	8.575  \\\tableline
6&	1226-1414&	0.800&	1.519&	6.780&	7.795  \\\tableline
7&	1414-1620&	0.761&	1.650&	5.583&	6.631  \\\tableline
8&	1620-1854&	0.749&	1.816&	4.421&	5.467  \\\tableline
9&	1854-2119&	0.678&  1.962&	3.621&	4.783  \\\tableline
10&	2119-2404&	0.712&	2.190&	2.940&	4.110  \\\tableline
11&	2404-2756&	0.656&	2.332&	2.265&	3.457  \\\tableline
\end{tabular}\label{tab:bknpower}
\end{centering}
\tablecomments{The parameters of XSPEC \textit{bknpower} model are shown. The low energy slope of the count spectrum is fixed at $\Gamma_1=-1$,
which corresponds to the Raleigh-Jeans tail of the blackbody emission with $F_\nu\propto \nu^2$. The host galaxy hydrogen column density
$N_H=0.750\times10^{22}{\rm cm}^{-2}$ with the metallicity $0.07$ of solar \citep{Wiersema2007} provides the best joint fit to all time slices with
$\chi^2=5463.0$ for ${\rm dof}=4836$. The galactic column is fixed at $N_H=9.4\times10^{20}{\rm cm}^{-2}$ following \citet{Campana2006}.}
\end{table*}

The results of spectral fitting are presented in Table~\ref{tab:compPS} for the blackbody spectrum Comptonized by thermal electrons as computed within a \textit{compPS} XSPEC model.
The optical depth of Compton scattering is fixed at $\tau_\sigma=1$ for scattering at the photosphere. The host galaxy hydrogen column
$N_H=1.090\times10^{22}{\rm cm}^{-2}$ provides the best joint fit to all time slices with $\chi^2=5358.9$ for ${\rm dof}=4836$ degrees of freedom.
The fit is slightly better for this model compared to the \textit{bknpower} model.
Our fit has a higher blackbody flux fraction at the peak ($40\%$) compared to $\sim15\%$ fraction in a fit by \citet{Campana2006},
who employed the solar metallicity of the absorber and used a sum of the power-law and the blackbody.
Table~\ref{tab:compPS} also summarizes application of the fireball model \citep{Peer:2007sd} to the blackbody emission component.
We can estimate the bulk Lorentz factor of the flow $\Gamma$ and the jet base radius knowing the temperature $T_{BB}$,
the observed isotropic flux $F_{BB}$, the distance to the source, and assuming a fiducial ratio $Y=10$ of the total fireball energy
to the energy emitted in X-rays/$\gamma$-rays. The obtained jet parameters are surprisingly consistent between different time slices, which might indicate that the fireball model captures
the physics of the event. The bulk Lorentz factor of $\Gamma\approx2.7 Y_{10}^{1/4}$ corresponds to a mildly relativistic jet, while the jet base radius is
$R_0=(1.0-1.7)\times10^{10}Y_{10}^{-3/2}$~cm.  The model is self-consistent for a wide range of $Y$ as the saturation radius $R_s=\Gamma R_0$ lies well within the photosphere
with the radius $R_{\rm ph}\sim5\times10^{12}$~cm. The jet launching region should be located at several BH gravitational radii. Assuming that the jet base radius is at $R_0=5\rg$,
we find the BH mass $M_{BH}=(1.3-2.3)\times10^4Y_{10}^{-3/2}M_\odot$, which places the BH into the intermediate mass category.
\citet{Peer:2007sd} argue that the ratio of energies is $Y_{10}\lesssim1$.
\begin{table*}
\scriptsize
\caption{Time-resolved spectroscopy of GRB060218 soft X-ray spectrum. Fitting with the blackbody spectrum Comptonized by thermal electrons.}\label{tab:compPS}
\begin{tabular}{ | p{7mm} | p{16mm}| p{16mm} | p{16mm} | p{19mm} | p{19mm} | p{19mm}| p{16mm}| p{20mm}|}
\tableline\tableline
  Number & Time period $t~[{\rm s}]$ & Photon temperature $T_0 [{\rm keV}]$ & Electron temperature $T_e [{\rm keV}]$  & Absorbed flux $F_{\rm abs} [10^{-9}{\rm erg~s}^{-1}{\rm cm}^{-2}]$ &
Unabsorbed source flux $F_{\rm unabs} [10^{-9}{\rm erg~s}^{-1}{\rm cm}^{-2}]$ & Blackbody source flux $F_{BB} [10^{-9}{\rm erg~s}^{-1}{\rm cm}^{-2}]$ & Lorentz factor $\times Y_{10}^{1/4},\Gamma$
& jet base radius $\times Y_{10}^{-3/2}, R_0 [10^{10}{\rm cm}]$\\
\tableline
1&	164-478&	0.1039&	262.3 &	4.586&	 6.61&	2.630& 2.667& 0.906\\\tableline
2&	478-691&	0.1052&	297.6&	7.159&	10.17&	4.033& 2.832& 1.091\\\tableline
3&	691-875&	0.0994&	258.8&	7.965&	11.49&	4.465& 2.803& 1.245\\\tableline
4&	875-1049&	0.1058&	230.2&	8.353&	12.29&	5.020& 2.898& 1.257\\\tableline
5&	1049-1226&	0.0967&	176.6&	7.424&	11.52&	4.763& 2.744& 1.493\\\tableline
6&	1226-1414&	0.1008&	145.8&	6.636&	10.84&	4.848& 2.753& 1.560\\\tableline
7&	1414-1620&	0.1119&	119.5&	5.517&	 9.56&	4.784& 2.816& 1.487\\\tableline
8&	1620-1854&	0.1116&	 93.1&	4.369&	 8.43&	4.686& 2.732& 1.733\\\tableline
9&	1854-2119&	0.1324&	 80.9&	3.614&	 7.16&	4.412& 2.879& 1.393\\\tableline
10&	2119-2404&	0.1489&	 65.9&	2.970&	 6.25&	4.284& 2.961& 1.275\\\tableline
11&	2404-2756&	0.1421&	 55.3& 	2.268&	 5.41&	3.963& 2.818& 1.485\\\tableline
\end{tabular}\label{tab:compPS}
\tablecomments{The parameters of the blackbody spectrum Comptonized by thermal electrons within a \textit{compPS} model are shown.
The optical depth is fixed at $\tau_\sigma=1$, which corresponds to Compton scattering at the photosphere.
We find the host galaxy hydrogen column density $N_H=1.090\times10^{22}{\rm cm}^{-2}$ in a self-consistent joint fit over all time slices
with $\chi^2=5358.9$ for ${\rm dof}=4836$ degrees of freedom. The Galactic column is fixed at $N_H=9.4\times10^{20}{\rm cm}^{-2}$ with solar metallicity following
\citet{Campana2006}. The quantity $Y_{10}$ is the ratio of the total fireball energy to the energy emitted in the X-rays in the units of $10$.
The last two columns are computed for $Y_{10}=1$ or $Y=10$ within the thermal fireball emission model described in \citet{Peer:2007sd}.
The jet base radius $R_0\sim10^{10}{\rm cm}$ corresponds to the BH mass $M_{BH}\sim10^{4}M_\odot$ assuming the jet is launched within several $\rg$ from the center.}
\end{table*}

\begin{figure*}[htbp]
    \plottwo{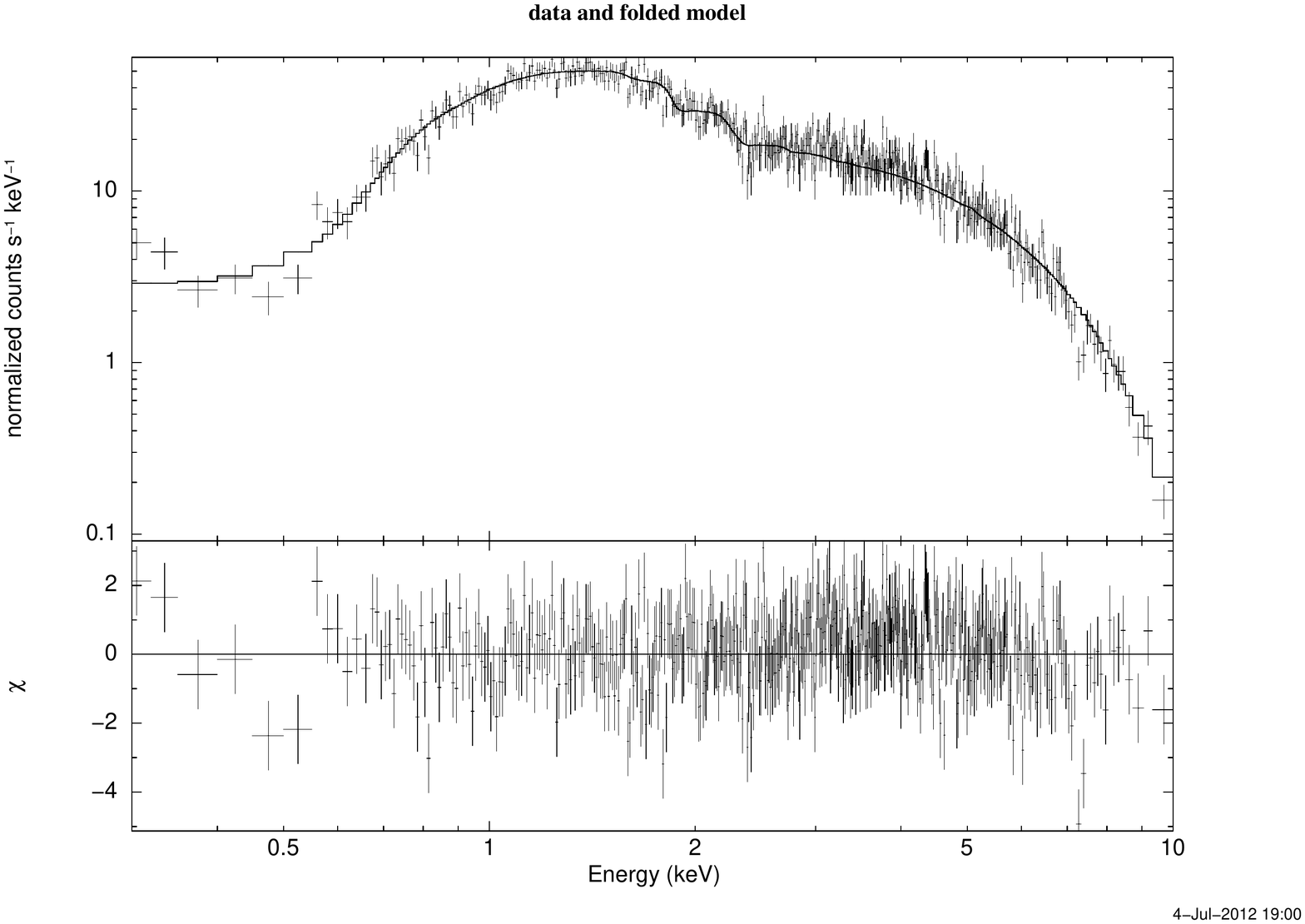}{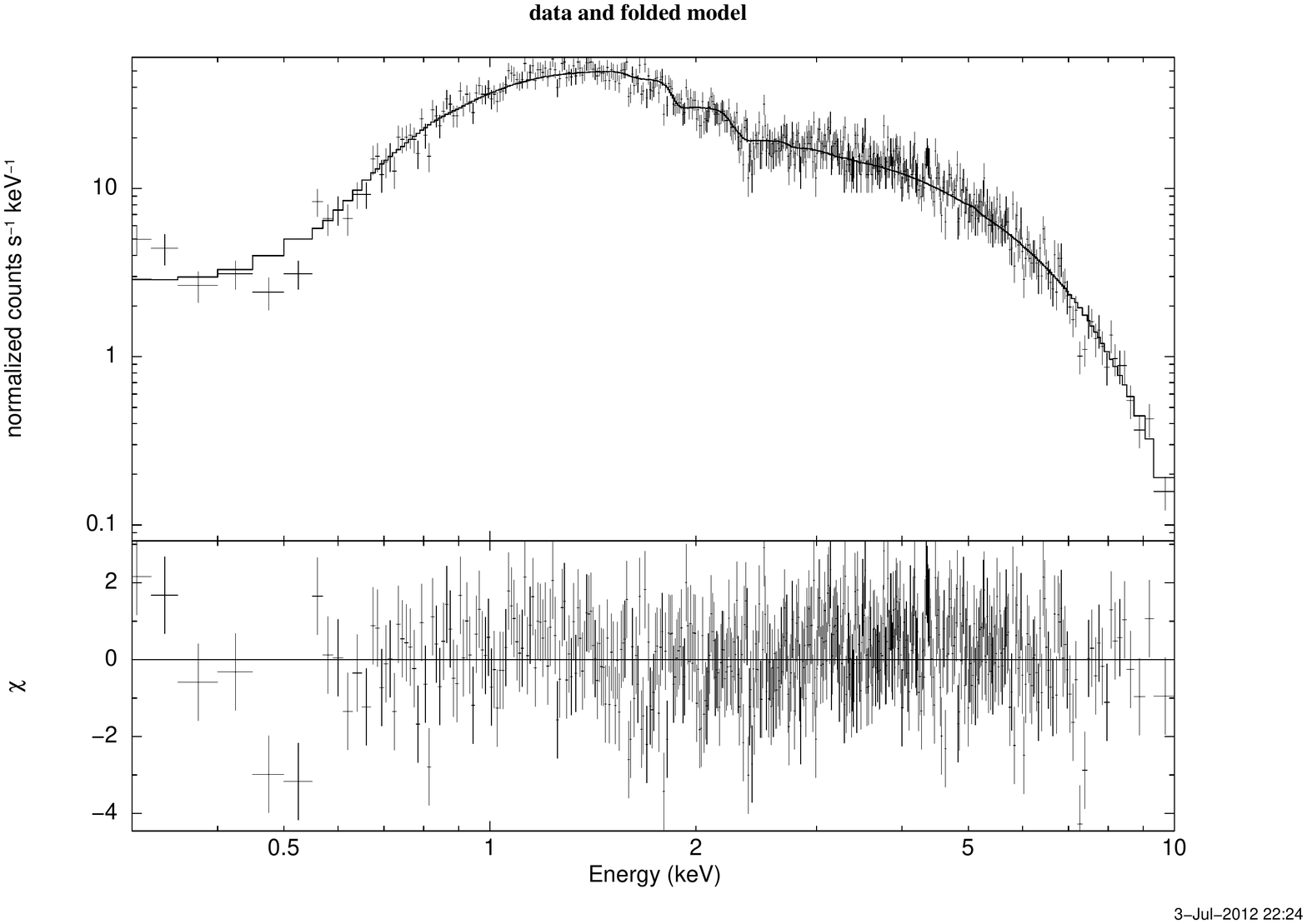}
    \caption{The parts of XSPEC spectral fits and the normalized residuals for the $4$-th time slice. This time slice corresponds to the highest soft X-ray flux.
The left panel shows the results for the \textit{bknpower} model, the right panel shows the results for the blackbody spectrum Comptonized by thermal electrons
within the \textit{compPS} model. The model parameters can be found in Tables~\ref{tab:bknpower} and \ref{tab:compPS}.}
    \label{fig:s4fit}
\end{figure*}

The reduced chi-squared $\chi^2/{\rm dof}=1.138$ for the joint \textit{bknpower} fit and $\chi^2/{\rm dof}=1.115$ for the joint \textit{compPS} fit are quite low despite good photon statistics.
Even better fits down to $\chi^2/{\rm dof}=1.0$ can be achieved, when the host galaxy hydrogen column density is allowed to vary between the time slices as in \citet{Butler:2007uy}.
However, we do not see a physical reason for $N_H$ to vary on the timescale of $10^3$~s, which corresponds to $3\times10^{13}$~cm distance, while the characteristic
star-formation size of the host galaxy is $R_{80}=0.55{\rm kpc}=10^{21}$~cm \citep{Svensson:2010ka}.  To test for potential discrepancies between the fits and the observations,
we depict on Figure~\ref{fig:s4fit} the parts of fits and the normalized residuals for the $4$-th time slice characterized by the highest soft X-ray flux.
The left panel shows the results for the \textit{bknpower} model, while the right panel shows the results for the blackbody spectrum Comptonized by thermal electrons.
The residuals do not systematically deviate from zero except slightly at the very low energies or for certain absorption features.
At the low energies \textit{Swift} XRT response might be uncertain or the line physics might alter the spectrum. The most prominent absorption feature is around $7.3$~keV.

\subsection{Prompt Optical/UV Emission}\label{subsec:promptUV}
The prompt optical/UV emission observed by the UVOT instrument onboard \textit{Swift} satellite was a matter of major disagreements between the former theoretical models.
The same blackbody component cannot produce both the optical and the X-ray emission \citep{Ghisellini:2007aa},
which is consistent with the physical picture within the tidal disruption scenario.
For an estimate we take the magnetic field in the photosphere to be in equipartition with the jet kinetic energy.
Then the Poynting energy flux is about $Y/2\sim5$ times larger than the total X-ray radiation flux.
Hence the magnetic field in the photosphere has the strength
\begin{equation}
B_{\rm ph}\sim10^5{\rm G}.
\end{equation}
The \textit{compPS} fit to the prompt X-ray spectrum shows that the non-thermal particles with the energies $\sim200$~keV exist in the photosphere.
The fact that the emission continues to the hard X-rays in BAT band reveals the existence of much more energetic particles with $\gamma\sim10-100$,
which readily emit synchrotron in the optical band. The synchrotron emission by an electron with a random Lorentz factor $\gamma\sim30$ peaks at $300$~nm for $10^5$~G magnetic field.
Since the effective synchrotron cross-section is much higher than the Thompson scattering cross-section, the optical photosphere is much above the X-ray photosphere.
In fact, very large jet particle density makes the optical photosphere coincide with the front of the jet propagating with a bulk Lorentz factor $\Gamma\sim3$ into the surrounding medium.
The jet front can propagate up to $R_{\rm front}=t_{\rm obs}\Gamma^2c$, where $t_{\rm obs}$ is the time since BAT trigger.
Continuous energy dissipation and particle acceleration at the front ensures there are enough energetic electrons to emit optical/UV synchrotron.
Note that the interactions with the interstellar medium (ISM) at the early times, which lead to a forward shock, may contribute less to the emission.
Low magnetic field at the jet front
\begin{equation}
B_{\rm front}\sim10^3{\rm G}
\end{equation}
at $t_{\rm obs}=3\times10^3$~s is compensated by higher Lorentz factors $\gamma\sim300$ of the electrons producing the optical/UV emission.

The idea of the jet front emission explains the steep $F_\nu\propto \nu^2$ spectrum and the achromatic flux growth at different UVOT frequencies till $\sim3\times10^4$~s.
The jet front is transparent to the hard UV and X-ray photons. Since the particles are subject to rapid cooling, the spectrum of re-accelerated particles
may substantially deviate from a power-law and resemble a very hot Maxwellian.
Then the resultant self-absorbed spectrum could be the Rayleigh-Jeans part of the blackbody $F_\nu\propto\nu^2$
as opposed to the absorbed synchrotron $F_\nu\propto\nu^{5/2}$ spectrum \citep{Rybicki1979}.
The observed temperature of the optical/UV radiation is about $10^6$~K \citep{Ghisellini:2007bb}, which violates the energetics of the source in a non-relativistic emission model with
the total emitted energy $E_{BB}\gtrsim10^{51}$~erg. In turn, the relativistic jet front emission does not violate the source energetics.
Relativistic Doppler boosting with $\Gamma\sim3$ reduces the emission temperature down to $\lesssim3\times10^5$~K, which readily leads to $E_{BB}\lesssim10^{50}$~erg.
The inferred large optical emission radius $10^{14-15}$~cm at $3\times10^3$~s is consistent with $R_{\rm front}$ for the mildly relativistic expanding shell.
A more detailed jet model for optical/UV emission by \citet{Ghisellini:2007aa} is similarly consistent with the source energetics.

\subsection{Temporal Analysis}\label{subsec:temporal}
\begin{figure}[htbp]
    \centering\plotone{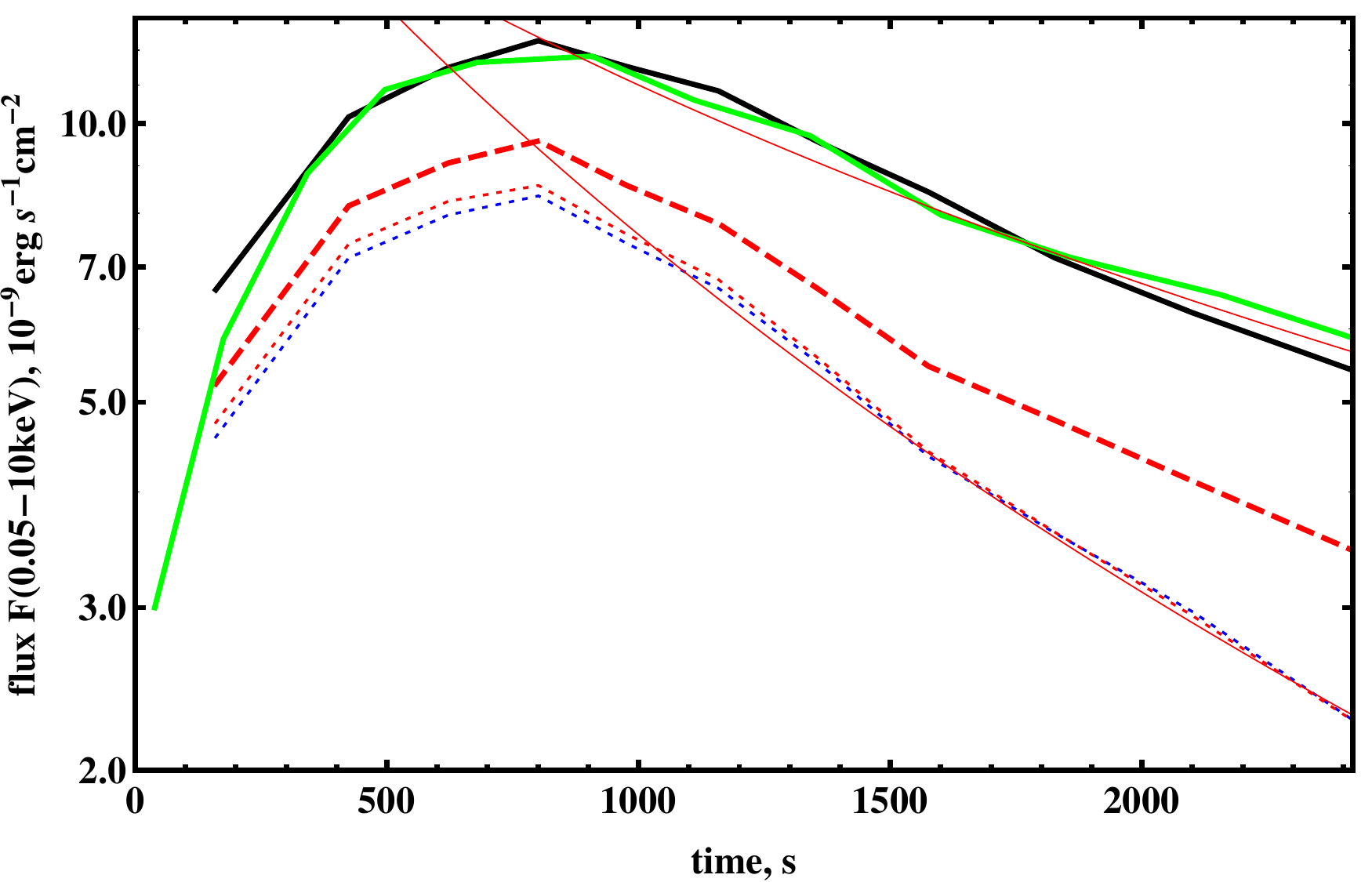}
    \caption{Soft X-ray lightcurve of GRB060218 fitted with the accretion rate temporal evolution $\dot{M}(t)$.
Shown are the unabsorbed source lightcurve for the \textit{compPS} model (dark solid) and for the \textit{bknpower} model (red dashed),
the observed absorbed lightcurve for the \textit{compPS} model (lower dotted) and for the \textit{bknpower} model (upper dotted), and
the scaled fallback accretion rate based on \citet{Laguna1993c} (light solid). The scaled accretion rate is offset by
$\Delta t=1810$~s, which the most bound debris takes to reach the BH. Two thin lines $(t+\Delta t)^{-3}$ and $(t+\Delta t)^{-5/3}$ are shown
to emphasize the asymptotic behavior of the absorbed flux and the unabsorbed flux, respectively.
While the observed flux decays as $F\propto t^{-3}$, the unabsorbed flux exhibits $F\propto t^{-5/3}$ late-time behavior in agreement with the tidal disruption scenario.}
    \label{fig:unlightcurve}
\end{figure}

The spectral modeling allows for proper determination of fluxes at different times to study the temporal behavior of the source.
Figure~\ref{fig:unlightcurve} shows the absorbed and the unabsorbed source fluxes as functions of time. The soft X-ray flux peaks at around $800$~s following \textit{Swift} BAT trigger.
We fit the lightcurve with the scaled accretion rate dependence on time $\dot{M}(t)$ for a tidal disruption of a MS star by a SMBH. Such a dependence for $\beta_T=1$ was taken
from \citet{Laguna1993c}. The best fit corresponds to a time delay of $\Delta t=1810$~s between the tidal disruption and \textit{Swift} BAT trigger.
The peak at $t_{\rm peak}\approx 2600$~s corresponds, for example, to the tidal disruption of a $m_{WD}=0.75M_\odot$ WD by a $1\times10^4M_\odot$ IMBH
(or a $m_{WD}=0.86M_\odot$ WD by a $2\times10^4M_\odot$ IMBH) at $\beta_T=1$. The tidal radius in this case is $16\rg$ (or $8\rg$).
A deeper encounter with $\beta_T>1$ may be needed to initiate a nuclear burning for an accompanying supernova,
but the dependence of critical $\beta_T$, for which the effective nuclear burning starts, on the WD and the BH masses is not known at present.
The encounter in question may indeed have very small pericenter distance with $\beta_T\gg1$ and not violate the peak time constraint.
The dependence of peak times on $\beta_T$ is found very weak in numerical simulations \citep{Colle:2012ad,Guillochon:2012la} as well as
in theoretical modeling \citep{Stone:2012pa}.
The accretion rate curve $\dot{M}(t)$ provides a good fit to the unabsorbed source flux. However, such fitting procedure is prone to multiple caveats.
First, the accretion disk onto the BH takes a finite amount of time $t_{\rm magn}\sim100t_{\rm orb}\sim250$~s to generate the strong poloidal magnetic field.
Despite the magnetic field generation time is much less than the characteristic dynamical time $t_{\rm magn}\ll t_{\rm dyn}\sim1000$~s, the precise fit is not expected.
Secondly, the transition from the accretion power $\dot{M}c^2$ to the emission power in a certain band is not trivial.
The jets are expected to be radiatively efficient \citep{Peer:2007sd}, and most of the emission, especially at the late times, falls into the $0.05-10$~keV band.
Only if the jet kinetic power is a constant fraction of the accretion power, then the soft X-ray luminosity is expected to approximately follow the accretion power.
Finally, the GR effects are not included in these estimates. \citet{Haas:2012ak} emphasized that the ultra-close encounters with small $R_P\sim {\rm several}\times\rg$
may lead to non-trivial temporal behavior of $\dot{M}$.

The estimated large BH mass and the small Lorentz factor may lead to a smooth lightcurve.
The lightcurves of typical GRBs are highly variable, which is attributed to fast variations near small BHs with masses $M_{BH}\sim10M_\odot$,
when the BHs launch shocks at different speeds. The variation timescale is correspondingly longer for the BHs with much larger masses $M_{BH}\sim10^4M_\odot$.
Those variations amplify, when the shocks moving at different high Lorentz factors collide.
The energy release and the radiated energy are relatively small, when the shocks with small $\Gamma$'s collide.
Lastly, the collisions of the slow-moving shocks are likely to happen inside the photosphere, so that any variations are smoothed.
Thus, a slow moving jet of GRB060218 is expected to vary on a very long timescale and the variations are expected to be small. The object should exhibit a smooth lightcurve.
\begin{figure}[htbp]
    \centering\plotone{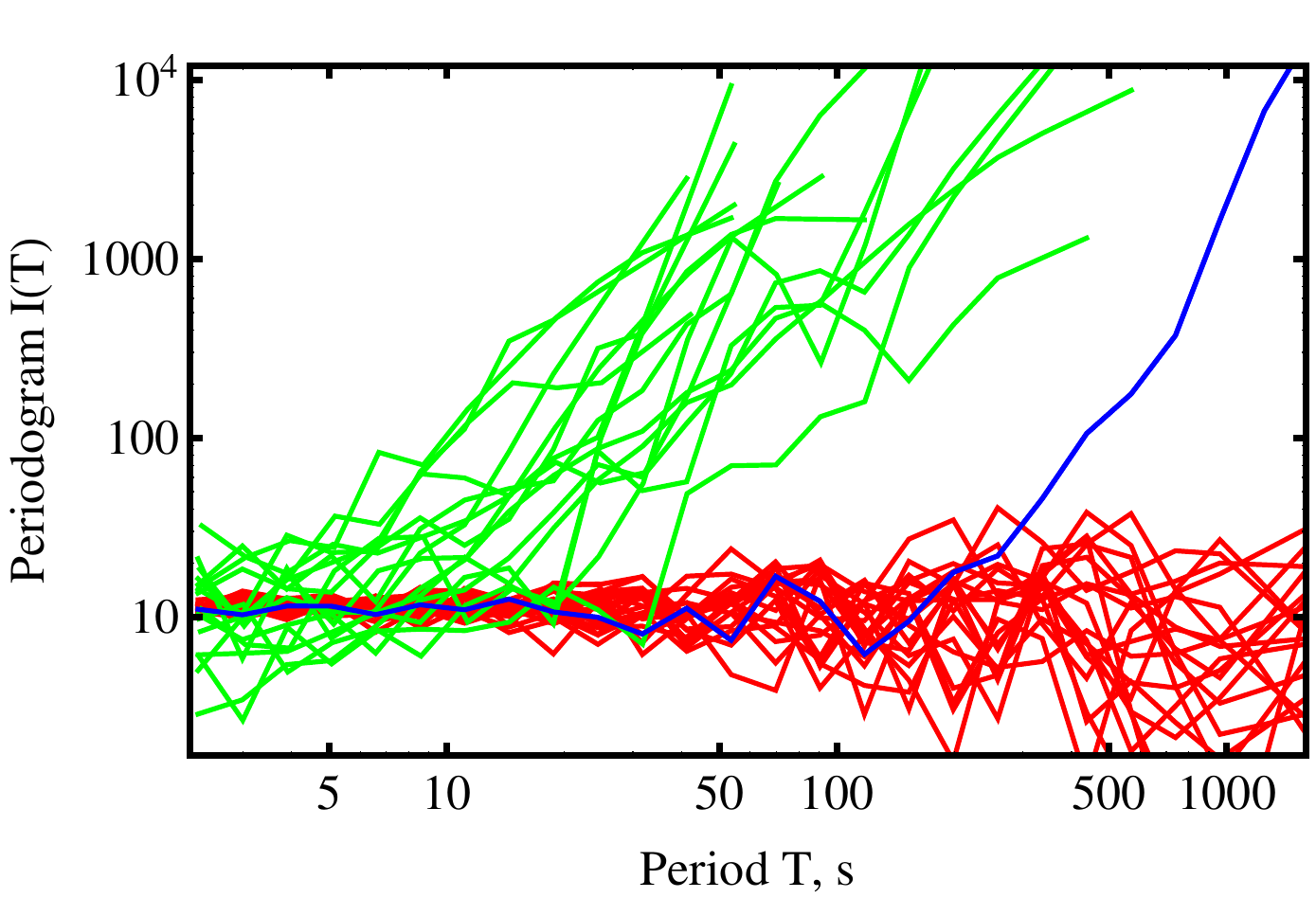}
    \caption{Log-smoothed periodograms $I(T)$ \citep{Papadakis93} of the lightcurves binned to $1$~s. The flat red curves depict simulated white noise,
    GRB060218 is shown by a dark/blue curve, while $16$ other GRBs are shown by light/green curves. The white noise has the same mean count rate and total counts as GRB060218.
    The other GRBs are selected to have the peak count rate between $142{\rm cts~s}^{-1}$ and $166{\rm cts~s}^{-1}$, while the peak count rate is $148{\rm cts~s}^{-1}$ for GRB060218.
    The break time $t_b$, at which the periodogram starts deviating substantially from the noise, corresponds to a characteristic system timescale.
    The break time for our event is $t_b\sim300$~s, which is consistent with the emission rise timescale in the tidal disruption.
    Other GRBs have substantially lower break times $t_b=5-40$~s, and the distribution of $t_b$ does not continue to $300$~s. This indicates a different origin of GRB060218.
    Log-smoothing is performed to $0.11$~dex.}
    \label{fig:periodogram}
\end{figure}
On Figure~\ref{fig:periodogram} we present the log-smoothed periodograms $I(T)$ \citep{Papadakis93} of GRB060218 and other lightcurves.
The dark (blue) rising curve is the periodogram of GRB060218 prompt X-rays. The dark (red) constant curves are the periodograms of $20$ white noise implementations
with the same mean count rates and total counts as GRB060218. The light (green) curves represent the selection from the online \textit{Swift} catalogue of $16$ GRBs,
whose peak photon count rates are the closest to the peak count rate $148{\rm cts~s}^{-1}$ of GRB060218.
The white noise implementations are flat as expected, while our candidate tidal disruption source rises above the noise at a very long break time of about $t_b=300$~s.
Other $16$ GRBs show much smaller break times of $5-40$~s. For some of the sources the break time coincides with the characteristic flux decay time, while most sources
exhibit variability on a shorter timescale. As predicted, the lightcurve of GRB060218 is smooth and shows no variability faster than the emission rise timescale.
This strengthens the case for the source being a slow dense jet from an IMBH.

\subsection{Steep Decay Phase and Afterglow}\label{subsec:afterglow}
The \textit{Swift} satellite was unable to observe the source from $2780$~s till $5900$~s counting from the BAT trigger.
The observed X-ray flux at $5900$~s is very low, which is seemingly inconsistent with $t^{-5/3}$ behavior of the accretion rate.
However, the blackbody source flux may still be large, while the observed flux may be a factor of $40$ lower due to the action of absorption and incomplete overlap with the XRT band.
The XRT instrument observes mostly Comptonized photons with its effective energy range $0.3-10$~keV.
The energy of the Comptonizing electrons goes down with time during the prompt phase, which
translates into the ratio of unabsorbed to absorbed fluxes rising from $1.4$ to $2.4$ by $11$-th time slice (see Table~\ref{tab:compPS}).
The ratio of fluxes keeps dropping till $10^4$~s in the steep decline phase. After that a much shallower decline $F\propto t^{-1.2}$ follows \citep{Campana2006} with
the spectrum consistent with a power-law $F_\nu\propto \nu^{-2.2}$ \citep{Soderberg:2006na}.
While the temporal decay of this late phase is consistent with the afterglow,
the spectrum is substantially softer than the typical afterglow spectrum emitted by the external shock \citep{Toma2007}.
The softer spectrum and the long afterglow may be explained by the late activity of the source \citep{Soderberg:2006na}.
This is an especially viable idea, since the jet kinetic luminosity is expected to stay high for a long period of time following a tidal disruption.

\begin{figure}[htbp]
    \plotone{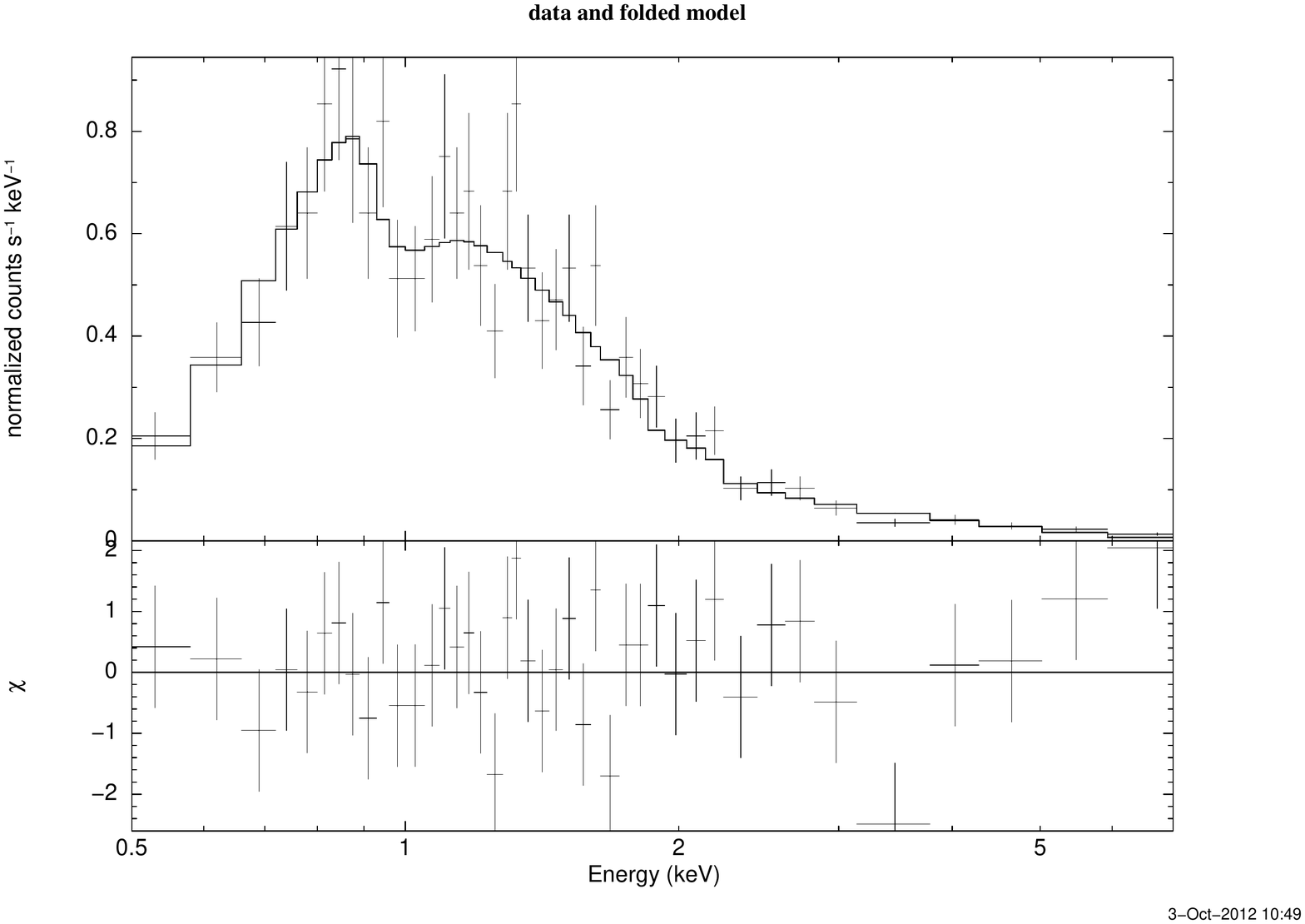}
    \caption{Best fit to the spectrum at the late times $t=(5950,7070)$~s with $\chi^2=37.3$ for ${\rm dof}=36$.
    The model consists of the blackbody emission Comptonized by thermal electrons absorbed by blueshifted oxygen and a powerlaw with a fixed slope $\Gamma_{pl}=2$ ($F_\nu\propto \nu^{-1}$).
    The sum of these two components is absorbed by the galactic $N_H=9.4\times10^{20}{\rm cm}^{-2}$ and $N_H=1.090\times10^{22}{\rm cm}^{-2}$ of the host galaxy
    with their respectively metallicities all fixed from the the early time modeling. The blueshift $z=-0.422$ corresponds to the bulk Lorentz factor $\Gamma=1.73$.
    The best-fitting cold oxygen column density is $N_O=1\times10^{18}{\rm cm}^{-2}$, which can readily be provided by a cooling jet exhaust.
    The blueshifted oxygen manifests as an absorption feature around $1$~keV.
    The unabsorbed source flux in $F_{\rm unabs}(0.05-10{\rm keV})=1.8\times10^{-9}{\rm erg~s}^{-1}{\rm cm}^{-2}$, which coincides with the extrapolation based on $t^{-5/3}$ law,
    while the heavily absorbed observed flux is $F_{\rm abs}=4.0\times10^{-11}{\rm erg~s}^{-1}{\rm cm}^{-2}$.}
    \label{fig:latespec}
\end{figure}

We model the steep decay spectrum within the time interval $t=(5950,7070)$~s  with a combination of the blackbody component Comptonized by thermal electrons
and the power-law emission from the external shock. We fix the absorption at the level determined with \textit{compPS} fitting of the prompt X-ray emission.
The fit allows for a large range of power-law slopes without a significant change in $\chi^2$. We fix the power-law slope at $\Gamma_{pl}=2$ ($F_\nu\propto \nu^{-1}$), which
corresponds to the afterglow of a typical GRB \citep{Soderberg:2006na}.
The absorbed combination of the \textit{compPS} and the \textit{powerlaw} models leads to $\chi^2=43.1$ for ${\rm dof}=36$.
As suggested by \citet{Butler:2007uy}, the spectrum contains lines, which become prominent at the late times.
Lines is the natural outcome of the atomic processes in a warm absorbing/emitting jet baryonic material.
The jet baryonic material consists of WD debris rich in oxygen and carbon and practically devoid of hydrogen.
The absorption by cooled down jet exhaust may become strong at the late times. We model such absorption by a blueshifted \textit{zvphabs} model, where we set
the abundance of a single chemical element to $1000$ and the abundances of the other elements to zero.
The model with a pure oxygen cold absorber converges to a blueshift $z=-0.422$, which corresponds to the bulk Lorentz factor $\Gamma=1.73$ consistently with the fits to the prompt spectrum.
The fit with $\chi^2=37.3$ for ${\rm dof}=36$ and the normalized residuals are depicted on Figure~\ref{fig:latespec}.
The fit can be marginally improved with the addition of iron elements. We do not present such modeling here for simplicity.
The addition of up to $30\%$ of carbon does not change the fit.
The best-fitting oxygen column density $N_O=1\times10^{18}{\rm cm}^{-2}$ can be readily provided by the cooling jet material.
Very soft blackbody has a temperature $T=0.052$~keV.
 The best-fitting flux is $F_{\rm unabs}(0.05-10{\rm keV})=1.8\times10^{-9}{\rm erg~s}^{-1}{\rm cm}^{-2}$, which directly traces $t^{-5/3}$ decay law from the time of peak flux,
while the observed absorbed flux $F_{\rm abs}=4.0\times10^{-11}{\rm erg~s}^{-1}{\rm cm}^{-2}$ is $40$ times lower.
The absorbed power-law contribution to the total source flux is $F_{\rm abs,pl}=2.5\times10^{-11}{\rm erg~s}^{-1}{\rm cm}^{-2}$,
which is consistent with the emission at later times being dominated by the power-law.
The application of the photospheric emission model \citep{Peer:2007sd} to determine the Lorentz factor and the jet base radius gives $\Gamma=1.45Y_{10}^{1/4}$
and $R_0=1.1\times10^{11}Y^{-3/2}$~cm, respectively.
A larger ratio of total to radiated energy $Y=4\times Y_{\rm prompt}$ makes the numbers consistent with the results from the prompt phase.
Since the blackbody is too soft to be directly observed,
there is a substantial degeneracy between its temperature and normalization, which leads to the unabsorbed flux being uncertain by a factor of $3$.
Taking this uncertainty into the account, the unabsorbed flux $F_{\rm unabs}$ is consistent with the $t^{-5/3}$ law even for larger $Y$.
We might not be able to compute a more self-consistent physical model of the steep decay phase unless
we incorporate the atomic physics of warm emitters/absorbers, which lies beyond the scope of the present paper.

The accretion rate law switches from $t^{-5/3}$ fallback dominated behavior to $t^{-4/3}$ behavior determined by disk viscous spreading \citep{Kumar:2008rn,Cannizzo2009,Cannizzo11}.
The time of this transition as estimated by \citet{Cannizzo11} is
\begin{equation}
t_{x1}=1.25\times10^{10}{\rm s}\left(\frac{M_\star}{M_\odot}\right)^{-3/2}\left(\frac{R_\star}{R_\odot}\right)^{3/2}\left(\frac{M_{BH}}{10^7M_\odot}\right)
\left(\frac{\alpha}{0.1}\right)\left(\frac{\eta}{2 \beta_T}\right)^{9/2},
\end{equation} where as before $\eta=2$ is the ratio of initial disk radius to pericenter radius.
The implicit assumption in \citet{Cannizzo11} is that all debris form a disk, which then viscously spreads.
However, the viscous time at the disk outer boundary, or the accretion time, is initially much lower than the time since disruption \citep{Ulmer:1999fg,Kumar:2008rn,Strubbe2009},
and the debris falls onto the BH as opposed to accumulating in a disk.
Let us take a proper account of fast accretion timescale and perform a refined estimate of the transition time $t_x$.
We make a one-zone approximation, where most mass and angular momentum of the disk are located near a single radius $R_{\rm disk}$.
Disk radius increases with time from the initial
$R_{\rm disk}=2R_P$. The disk mass $M_{\rm disk}$ also grows with time. Relatively little angular momentum transfer happens between the disk and the BH.
The angular momentum of debris
\begin{equation}\label{eq:Ldiskin}
L_{\rm disk}=M_\star\sqrt{G M_{BH}R_P/2}=\rm const
\end{equation} becomes the preserved angular momentum of the disk after most of debris fall onto the BH. We took the debris mass to be $M_\star/2$. The disk angular momentum
is expressed through current mass and radius of the disk as
\begin{equation}\label{eq:Ldisk}
L_{\rm disk}=M_{\rm disk}\sqrt{G M_{BH}R_{\rm disk}}.
\end{equation} The viscous time of the disk or the accretion time is
\begin{equation}\label{eq:tvisc}
t_{\rm visc}=\frac{R_{\rm disk}^{3/2}}{\sqrt{\alpha G M_{BH}}}
\end{equation} for disk thickness $H=R$.
The infalling debris stay within the disk for only $t_{\rm visc}$, thus the disk mass is
\begin{equation}\label{eq:Mdisk}
M_{\rm disk}=\dot{M}_{fb}t_{\rm visc},
\end{equation} where the fallback rate $\dot{M}_{fb}$ is given by formula (\ref{eq:Mdot}).
It can be seen from above equations that $t_{\rm visc}\propto t^{5/4}$. While initially the viscous time is small $t_{\rm visc}<t$ as noted by \citet{Ulmer:1999fg,Strubbe2009},
the equality is achieved later at the transition time $t_x$. Solving the equations (\ref{eq:Ldiskin}-\ref{eq:Mdisk}) we find another estimate for the transition time
\begin{equation}
t_{x2}=\frac{2\sqrt{2\alpha G M_{BH}}t_{fb}^2}{27R_P^{3/2}},
\end{equation} where the fallback time $t_{fb}\approx2700$~s is inferred from the observed lightcurve.
At late times $t>t_{x2}$ the viscous time becomes longer than the time since disruption.
Then the fallback-dominated behavior gives way to spreading disk behavior, and the disk mass becomes larger than the mass of debris yet to fall back onto the disk.

Taking a WD with a mass $0.86(0.75)M_\odot$ and a radius $R_\star=7.1(8.1)\times10^8$~cm disrupted by a $2.0(1.0)\times10^4M_\odot$ BH we estimate the transition time to be
$t_{x1}=2.6(1.6)\times10^4$~s and $t_{x2}=5.6(3.0)\times10^4$~s for $\beta_T=1$. The estimated transition time $t_x$ crudely agrees to observations. Shallow disk spreading behavior
takes over around the transition to the afterglow phase. The temporal index of the power-law flux decay $-1.2\pm0.1$
during the afterglow \citep{Soderberg:2006na} is consistent with $-1.33$ index of the mass accretion rate at the late times.
Despite estimated transition time crudely agrees to observations, our one-zone calculation may not offer an ultimate answer for $t_x$.
A self-similar solutions of disk spreading behavior by \citet{Lynden-Bell:1974xp} show a small fraction of mass carrying angular momentum to a large radius,
while the bulk of disk mass is concentrated at a smaller distance from the central object serving as a reservoir of the infalling gas.
Correspondingly, a refined transition time estimate may show a longer $t_x$.  Another important effect is cooling of the outer disk,
which leads to its collapse onto an equatorial plane accompanied by a dramatic increases of the viscous time.
A detailed computation is beyond the scope of the present manuscript. We leave it for future research.

\subsection{Associated Supernova and Host Galaxy}\label{subsec:superno}
Another challenge to the WD/IMBH tidal disruption model is the associated supernova.
As discussed in \S~\ref{sec:grbobs}, the supernova ejecta mass $M_{\rm ej}\sim(1-2)M_\odot$ is consistent with the high end of the WD mass distribution.
The unbound mass fraction is influenced by the interplay between the energy spread during the tidal disruption and the energy release in the supernova explosion.
The outflow velocity of tidal disruption debris is $20\times10^3{\rm km~s}^{-1}$ for the WD and the BH masses estimated above,
which is comparable to the typical outflow velocity in a Type Ia supernova $(10-20)\times10^3{\rm km~s}^{-1}$ \citep{Wang:2009ly}.
Thus, the supernova SN2006aj with the energy release comparable to that of a Type Ia supernova \citep{Khokhlov:1993al}
can unbind most of the WD material, which then contributes to the supernova ejecta mass. Only a small fraction of debris accretes back onto the BH.
The nebular phase of SN2006aj exhibits strong oxygen emission \citep{Mazzali:2007as}, which is consistent with the ignition of a carbon/oxygen WD.

A supernova explosion accompanying a tidal disruption is expected to be heavily asymmetric.
In particular, there could be a velocity shift of supernova lines with respect to the velocities of the host galaxy.
Yet, the velocities and the redshift $z\approx0.033$ were only estimated for the host galaxy \citep{Mirabal:2006oa,Modjaz:2006ua} and no measurements of supernova velocities exist.

\begin{figure}[htbp]
    \centering\plotone{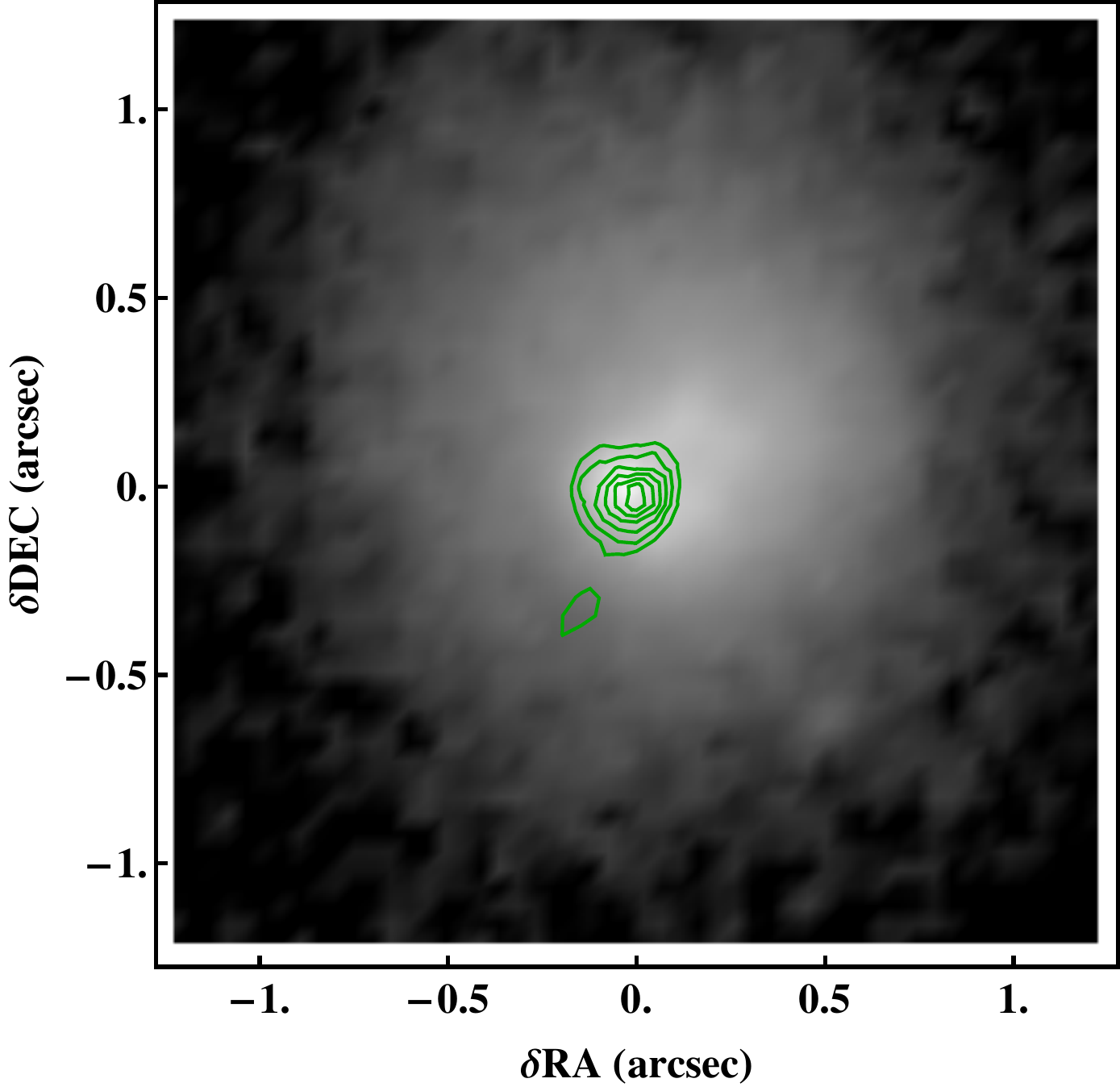}
    \caption{Position of SN2006aj supernova (green contours) within the host galaxy (grayscale intensity plot) from \textit{Hubble} ACS observations in red F814W band.
The supernova position is consistent with the center of the host. The supernova image is computed as the subtraction of the late time (26 Nov 2006)
image from the early time (18 Jul 2006) image. The late time image without supernova subtraction is taken to represent the host galaxy, which has irregular morphology.
The subtraction is not performed to avoid the oversubtraction of the galactic nucleus were the supernova to happen at the very center.}
    \label{fig:position}
\end{figure}

Since the IMBHs are expected to reside in the nuclei of dwarf galaxies, the other major test is the coincidence of the supernova position with the
center of the host galaxy. A star-formation radius for GRB060218 host is $R_{80}=0.55$~kpc \citep{Svensson:2010ka}, which corresponds to a half-light radius $R_{50}\approx 0.45$~kpc.
The supernova and the host galaxy were observed with \textit{Hubble} ACS instrument at different epochs.
Figure~\ref{fig:position} shows the position of the supernova (green contours) relative to the intensity plot of the host galaxy at a late time.
The host galaxy has a somewhat irregular morphology as can be seen on images by \citet{Misra:2011jg}, which complicates finding its center.
The uncertainty of center determination is about $dR\lesssim0.1{\rm arcsec}\approx70$~pc.
The supernova position appears to be consistent with the center to within $0.1{\rm arcsec}$.
The ratio of the uncertainty $dR$ to the half-light radius is $dR/R_{50}<0.15$, which is much less than a normalized offset $dR/R_{50}=0.98$
found for a representative set of long GRBs \citep{Bloom:2002of}. It is very rare for a supernova to happen that close to a galactic center at random.

The stellar mass of the host galaxy is estimated to be $M_{\rm st}\sim10^{7.2}M_\odot$ \citep{Ferrero:2007jd}.
Observations suggest that the bulge mass to the total stellar mass ratio ($B/T$) of galaxies is within $15-100\%$ range \citep{Schramm:2012ak}
and  $B/T=15\%$ for the Milky Way \citep{McMillan:2011dq}.
Then, extrapolating $M_{\rm bulge}-M_{BH}$ relation \citep{Magorrian:1998aj,Marconi:2003le}
\begin{equation}
M_{BH}/M_{\rm bulge}\sim0.002,
\end{equation} to low masses, we get the central BH mass to be $M_{BH}\sim(0.5-3)\times10^4M_\odot$.
This estimate, though much less certain, is consistent with two previous estimates based on the jet thermal emission and the lightcurve peak time.

\subsection{Accretion Flow and Jet Energetics}\label{subsec:energetics}
Let us compute the energetics of the jet for the estimated BH mass.
Assuming the supernova does not unbind the gas, the accretion rate onto the BH peaks at
\begin{equation}\label{eq:Mdotval}
\dot{M}\approx 2\times10^3M_\odot {\rm yr}^{-1},
\end{equation} which corresponds to the accretion power $\dot{M}c^2\approx1\times10^{50}{\rm erg~s}^{-1}$. Then the Blandford-Znajek jet power is
$P_{BZ}\sim0.01\dot{M}c^2\sim1\times10^{48}{\rm erg~s}^{-1}$ \citep{McKinney2005} for the low efficiency associated with simulations initiated with the weak magnetic field.
The isotropic equivalent radiation power for the opening angle $\theta=0.5$ and the ratio $Y=10$ is $P_{\rm rad,iso}\sim2\times10^{48}{\rm erg~s}^{-1}$.
This is substantially higher than the observed peak isotropic luminosity $P_{\rm obs,iso}=3\times10^{46}{\rm erg~s}^{-1}$.
The discrepancy can be partially explained by a small fraction of a WD material falling back onto the BH, since the supernova may unbind most of the gas.
Other potential explanations include GRB060218 happening substantially off-axis, the WD being mostly unbound in an encounter with a spinning BH \citep{Haas:2012ak},
or the source failing to launch a powerful jet. The latter is especially viable, since the source has a very limited amount of time to generate the ordered magnetic field.
A poloidal magnetic field $10$ times weaker than the equipartition value given by the equation~(\ref{eq:Beq}) leads to consistent jet energetics.
The disk may launch a mildly relativistic outflow by itself via the Blandford-Payne mechanism \citep{Blandford1982}.
An off-axis GRB is inconsistent with the absence of a jet break \citep{Soderberg:2006na}.

The X-ray emission is dominated by the super-Eddington jet at the early times, but the accretion disk may be brighter at the late times.
Our disruption models have the peak accretion rate of about $\dot{M}_{\rm peak}\approx2\times10^3M_\odot{\rm yr}^{-1}$ at the time $2600$~s after the disruption
(neglecting the effect of blowing the material away by a supernova). Then assuming $t^{-5/3}$ law we find that the accretion rate is down to the Eddington value for
a standard $\varepsilon=0.1$ efficiency at $t_{\rm Edd}\sim1$~yr. As described in \S~\ref{subsec:afterglow}, the accretion rate is expected to follow a shallow $t^{-4/3}$
slope after about $t_x\sim2\times10^4$~s due to the viscous evolution in the disk.
Yet, as we will describe below in \S~\ref{subsec:MS_IMBH}, when the outer disk boundary diffuses out to be much larger than $\sim20\rg$,
then the energy transfer between the inner and the outer disk kicks in.
The energy from the inner flow unbinds the outer material and the accretion rate onto the BH dramatically decreases.
\citet{Kumar:2008rn} derived the temporal dependence of the accretion rate in the RIAF with the energy transfer/convection to be $t^{-4(1+s)/3}$, where the parameter $s$ controls
the convection efficiency. The value of $s$ varies from $s=0$ for no energy transfer up to $s=1$ for the maximum energy transfer power.
Then the accretion rate may decrease as steeply as $t^{-8/3}$ in the RIAF phase. We will discuss the RIAFs with convection in more detail in \S~\ref{subsec:MS_IMBH}.
The onset of RIAF with the energy transfer may explain the steeper XRT flux dependence $F\propto t^{-1.55}$ at times $t>2\times10^5$~s, accompanied by
a nearly constant hardness ratio. Therefore, the accretion rate may fall below the Eddington rate at times $t\ll1$~yr.

Yet, the Eddington luminosity for a $2\times10^{4}M_\odot$ BH corresponds to a flux
\begin{equation}\label{eq:FEdd}
F_{\rm Edd}\approx1\times10^{-12}{\rm erg~s}^{-1}{\rm cm}^{-2},
\end{equation}
 which is about a factor of $100$ higher
than the late-time flux observed by the XRT.  This is not a contradiction, however, as at times much before $1$~yr the accretion disk might not be visible
at all due to obscuration by the outflowing debris. As discussed in our previous paper \citep{Haas:2012ak},
the outflowing debris consisting of carbon and oxygen have high absorption cross-section
$\sigma\approx2\times10^{-20}{\rm cm}^{-2}$ across $0.5-10$~keV band. The observed matter velocity in SN2006aj is about $1.5\times10^4{\rm km~s}^{-1}$ \citep{Mazzali:2006na}.
A tidal disruption may scatter matter in particular directions leaving huge voids, though which the inner disk could be observed.
A tidal disruption of an object with the size comparable to the tidal radius rather tends to scatter the debris over the large solid angle \citep{Haas:2012ak}.
In addition, the supernova, whose energy release is comparable to the energy release in a tidal disruption, should scatter the debris more uniformly in all directions.
Following \citet{Haas:2012ak} we find that the absorption optical depth unity $\tau=1$ is achieved around $t\sim3$~yrs, at which time the source is expected to rebrighten.
As the accretion disk should have a substantially sub-Eddington accretion rate by $3$~yrs, the emitted flux levels $F_X\lesssim10^{-13}{\rm erg~s}^{-1}{\rm cm}^{-2}$.
The accretion flow with the sub-Eddington accretion rate settles into a thin disk \citep{Shakura1973}, which has the inner temperature about
\begin{equation}
T_{\rm in}\approx\left(\frac{G M_{BH} \dot{M}}{\sigma_F R_{ISCO}^3}\right)^{1/4}.
\end{equation} The temperature is $T_{\rm in}\approx0.15$~keV for the accretion rate equal to $10\%$ of the Eddington value and the BH mass $2\times10^4M_\odot$.
Very soft blackbody spectrum absorbed by the galactic column and the host galaxy column leads to the ratio $F_{\rm unabs}/F_{\rm abs}\sim10$
of the model flux to the observed absorbed flux and the observed flux $F_{\rm abs}\sim 10^{-14}{\rm erg~s}^{-1}{\rm cm}^{-2}$, which might not be detectable.
If rebrightening occurs earlier, while the disk is still radiation-dominated and geometrically thick, then the expected observed temperature of the slim disk spectrum
is $T_{\rm slim}\approx0.6$~keV \citep{Haas:2012ak}. Most of slim disk emission would lie in X-rays with only under $1\%$ contribution by optical/UV at photon energies below $10$~eV.
A possibility that the disk dominates the emission starting as early as $10^4$~s, at which point the source flux equals the Eddington flux of a hypothesized IMBH,
is unrealistic. The observed X-ray flux $F_X\sim10^{-11}{\rm erg~s}^{-1}{\rm cm}^{-2}$ and the dereddened optical/UV fluxes $F_{\rm opt/UV}\sim10^{-11}{\rm erg~s}^{-1}{\rm cm}^{-2}$
\citep{Ghisellini:2007bb} are substantially super-Eddington according to equation (\ref{eq:FEdd}).


\section{EVENT RATES}\label{sec:rates}
The IMBHs are thought to reside either in the GCs or in the dwarf galaxies.
As we discussed in the earlier work \citep{Haas:2012ak}, the disruption rate of stars by IMBHs in GCs is very uncertain.
\citet{Baumgardt2004} predicts that for a GC with a $10^3M_\odot$ central IMBH, the optimistic disruption rate of stars is $10^{-7}{\rm yr}^{-1}$ per GC, while $15\%$
of all disruptions are those of WDs.
Then, following \citet{McLaughlin:1999ty} we theoretically estimate the space density of the GCs to be
\begin{equation}
n_{GC,{\rm th}}\approx34{\rm Mpc}^{-3},
\end{equation} which leads to the rate \citep{Haas:2012ak} $R_{WD-IMBH}\sim500{\rm yr}^{-1}{\rm Gpc}^{-3}$ of the WD disruptions by the IMBHs.
Observational constraints on the GC population suggest a slightly lower space density of the GCs \citep{Brodie:2006ac,Ramirez-Ruiz:2009er}
\begin{equation}
n_{GC,{\rm obs}}\approx4{\rm Mpc}^{-3},
\end{equation} so that the rate estimate should be revised down to
\begin{equation}\label{eqn:rate_est_GC}
 R_{WD-IMBH}\sim 50{\rm yr}^{-1}{\rm Gpc}^{-3}.
\end{equation}

The disruption rate of stars in dwarf galaxies was estimated to be very high.
If $M-\sigma$ \citep{Gebhardt2000} and $M_{BH}-M_{\rm bulge}$ \citep{Magorrian:1998aj,Marconi:2003le} relations hold down to the low BH masses,
then the predicted tidal disruption rate is higher for the dwarf galaxies compared to the other galaxies \citep{Wang:2004pl}.
Dependent on the radial stellar profile the disruption rate in a dwarf galaxy with a $10^4M_\odot$ BH varies from $R_{IMBH}=10^{-5}{\rm yr}^{-1}$
per galaxy up to $R_{IMBH}=10^{-1}{\rm yr}^{-1}$ per galaxy. For an estimate we take the rate of $R_{IMBH}=10^{-3}{\rm yr}^{-1}$ per galaxy.
Note that during a tidal disruption some fraction $f$ of a WD accretes onto a BH. Then the mass of an IMBH grows at a rate
\begin{equation}
\frac{dM_{BH}}{dt}=f R_{IMBH}m_{WD}.
\end{equation} Taking a large fiducial fraction $f=0.5$ we find that the BH mass doubles every $t_d=2\times10^8$ years solely due to tidal disruptions of WDs,
where we assumed that the same $R_{WD-IMBH}/R_{IMBH}=15\%$
of the disrupted stars are the WDs. This doubling is consistent with the estimated age of GRB060218 host galaxy $200$~Myr \citep{Ferrero:2007jd}.
As we will show in the next subsection, an IMBH swallows only a small fraction of a MS star, so that the IMBH mass growth rate due to the MS stars may not be dominant.

According to observations \citep{Ferguson:1991ge} the number density of dwarf \textit{d} galaxies with V magnitude $M_V>-15.5$ is about the number density
of large elliptical \textit{E} and spiral \textit{S} galaxies. Thus, the space density of dwarf galaxies is $n_{\rm dwarf}\sim10^{-2}{\rm Mpc}^{-3}$.
Then, assuming that every dwarf galaxy has a small enough IMBH, we arrive at a rate estimate
\begin{equation}\label{eqn:rate_est_DW}
R_{WD-IMBH}\sim1500{\rm yr}^{-1}{\rm Gpc}^{-3},
\end{equation} which is much larger than the rate in GCs.
One event per year for the field of view (FOV) $10\%$ of the sky would happen as close as $200Mpc$.
Such disruptions should be detected by \textit{Swift} BAT instrument with FOV of approximately $10\%$ of the sky.
Collimation into a wide jetted outflow may reduce the rate down to one event per several years consistently with a single GRB060218 source at $143$~Mpc distance.

Since the WD disruptions by the IMBHs produce supernovae, we can compare the rate of those with the total supernova rates.
\citet{Li:2011sn} find the local rate of supernovae Type Ia to be
\begin{equation}
R_{SN,Ia}\approx3\times10^4{\rm yr}^{-1}{\rm Gpc}^{-3},
\end{equation} and the rate of supernovae Type Ib/c to be $R_{SN,Ibc}\approx3\times10^4{\rm yr}^{-1}{\rm Gpc}^{-3}$.
In an optimistic estimate, if a supernova is produced in every tidal disruption of a WD, then one disruption-induced supernova happens per $20$ Type Ia supernovae.
A distinctive feature of disruption-induced supernovae is their location in the galactic nuclei substantially close to the photometric centers of their dwarf host galaxies.
This rate equals to the rate of calcium-rich gap transients, whose properties may be explained by the underluminous explosions of the WDs \citep{Kasliwal:2012oq}.
 However, those transients occur very far from the galactic nuclei.

\subsection{Disruptions of MS Stars by IMBHs}\label{subsec:MS_IMBH}
In the present paper we consider the radiative signatures of the WD disruptions by the IMBHs.
However, the disruptions rates of the MS stars by the IMBHs are several times higher.
Then the question is whether these more frequent events produce equally distinct and observable signatures.
In this subsection we show that MS star/IMBH disruptions are quite different: they are much longer and much fainter.
Despite lower rates, it might be easier to observe the WD/IMBH disruptions compared to the MS star/IMBH disruptions.

The fallback timescale is about $t_{\rm fb}\sim20$~days for a disruption of a $1M_\odot$ MS star with a solar radius by a
$10^4M_\odot$ IMBH with a pericenter radius equal to the tidal radius $R_P=R_T$.
 The peak fallback rate is $\dot{M}_{\rm peak}\sim5M_\odot{\rm yr}^{-1}$.
While a tidal disruption radius is $R_T\sim20\rg$ for a WD, a MS star with the solar mass and the solar radius has a much larger $R_T\sim1000\rg$.
Since the peak fallback accretion rate $\dot{M}_{\rm peak}\sim5M_\odot{\rm yr}^{-1}$ is much above the Eddington rate $\dot{M}_{\rm Edd}\sim2\times10^{-4}M_\odot{\rm yr}^{-1}$,
the RIAF settles from the outer disk radius $R_{\rm disk}\approx 2R_P$ down to the BH. A RIAF with an outer radius greater than about $20\rg$ settles
into the convection dominated accretion flow (CDAF) \citep{Narayan:2000tr,Quataert:2000er}, where the energy transport between the inner and the outer flow starts to play a role.
Unlike the fallback disks following the WD disruptions, the disks following the disruptions of the MS stars always exist in CDAF state.
A flow with convection has a shallow density profile $\rho\propto r^{-\beta}$ down to the inner radius $R_{\rm in}\sim20\rg$ \citep{Abramowicz:2002hg}.
Only a small fraction of available matter accretes in CDAF state. The sustained BH accretion rate is
\begin{equation}\label{eq:MdotCDAF}
\dot{M}=\dot{M}_{\rm fb}(R_{\rm disk}/R_{\rm in})^{1.5-\beta}.
\end{equation}
The density slope $\beta=1.5-s=0.5-1.0$ was found in the numerical simulations (see \citealt{Yuan:2012lp} for the review).
The value $\beta=0.8-0.9$ was estimated by \citet{Shcherbakov:2012appl} for Sgr A*.
The flow settles on a viscous timescale $t_{\rm visc}\sim1$~day, which is shorter than $t_{\rm fb}\sim20$~days.
Thus, the peak accretion rate of the CDAF is given by the formula~(\ref{eq:MdotCDAF}) with $\dot{M}_{\rm fb}=\dot{M}_{\rm fb, peak}$.
The peak accretion rate onto the BH can be a factor of $10-100$ lower than the peak fallback rate
\begin{equation}
\dot{M}_{\rm peak}\sim(0.01-0.1)\dot{M}_{\rm fb, peak}.
\end{equation}

Let us make a fiducial estimate for a BH mass $10^4M_\odot$ and a disruption of a Sun-like star for $R_P=R_T$ assuming density slope $\beta=0.85$.
In this case the accretion rate is lowered by a factor of $\approx23$ due to the action of convection according to the formula~(\ref{eq:MdotCDAF}), and the resultant jet power is
$L_{\rm kin}=0.01\dot{M_{\rm peak}}c^2\sim10^{44}{\rm erg~s}^{-1}$ with the same fiducial efficiency $P_0\sim0.01$ used in estimates for the WD disruptions.
A jet with such low power has the photospheric radius smaller than the saturation radius.
The radiation decouples from matter before the jet can accelerate, and the powerful photospheric emission is not expected.
The photospheric radius corresponds to the bulk Lorentz factor $\Gamma\approx2$. If the source is beamed towards us, then the observed temperature is
$T_{\rm ob}\approx0.5$~keV. The radiated blackbody luminosity is $L_{BB}\sim1\times10^{43}{\rm erg~s}^{-1}$, which corresponds to the distance $d\sim10$~Mpc to match
the peak flux level $F_{\rm peak}\approx10^{-8}{\rm erg~s}^{-1}{\rm cm}^{-2}$ of GRB060218.
The absorption by the cooled down jet exhaust or the dense wind, which consist mostly of hydrogen and helium, may substantially lower the X-ray luminosity and turn the source into
an optical/UV transient similar to the disruption of a MS star by a SMBH \citep{Strubbe2009,Strubbe:2011oa}.
The optical/UV transients lasting for $20$~days and releasing up to $10^{49}{\rm erg~s}^{-1}$ may be confused with variability of the AGNs.
According to the standard theory \citep{Cannizzo2009,Cannizzo11} at $t_x\sim100$~days the source should switch from the fallback-dominated $t^{-5/3}$ behavior
to the disk spreading behavior $t^{-4(1+s)/3}\propto t^{-2.2}$. Yet, the timescale $t_x$ was derived for the adiabatic RIAFs.
The temporal behavior of radiatively inefficient fallback accretion disks with convection were not explored to our knowledge
and should be investigated in the future work. If absorption is inefficient, then one transient per year is expected at a distance $d\sim100$~Mpc
with the peak flux level of about $F_{\rm peak}\approx10^{-10}{\rm erg~s}^{-1}{\rm cm}^{-2}$. Searches within the existing and future X-ray and optical surveys
should identify such sources and constrain the rates of tidal disruptions of various kinds. ROSAT All-sky survey search revealed several nearby tidal disruption candidates
with redshifts $z\lesssim0.1$ \citep{Komossa:2002pr}.
Some of those candidates, e.g. RX J1242-1119, cannot be easily matched with a large host galaxy \citep{Komossa:1999pa,Komossa2004}.
The absence of suitable X-ray/optical transients in the data might not immediately invalidate high tidal disruption rates.
For example, if the jet power in the disruptions of the MS stars is as weak as in GRB060218,
then the X-ray/optical flux from such disruptions is expected to be up to $100$ times lower.
The detailed analysis of the tidal disruptions of the MS stars is beyond the scope of the present manuscript.

A candidate source for a stellar disruption by an IMBH is a source in a GC NGC1399 explored by \citet{Irwin:2010jc}.
Observed at a distance $20$~Mpc the source exhibited X-ray luminosity of $2\times10^{39}{\rm erg~s}^{-1}$ and luminosity of a few $\times10^{36}{\rm erg~s}^{-1}$
in oxygen and nitrogen optical lines.  These properties were found consistent with a disruption of a horizontal branch star by a $50-100M_\odot$
BH \citep{Clausen:2012af}.

\section{DISCUSSION AND CONCLUSIONS}\label{sec:disc}
In this paper we explore the theory of the WD tidal disruptions by the IMBHs, an understudied, but a very promising type of an encounter.
The high rate of the WD tidal disruptions in dwarf galaxies warrants the search for candidates among already observed objects.
We identify GRB060218 with the accompanying supernova SN2006aj as a promising candidate and model its temporal/spectral properties and the properties of the host galaxy.
GRB060218 was explored previously within the supernova shock breakout model and the model, where a jet is launched by a newborn BH or a neutron star.
We find the tidal disruption model to be a viable alternative, which more naturally explains some features of the candidate.
In this section we briefly discuss the application of the tidal disruption model to GRB060218/SN2006aj source, and compare different models of the source.

The tidal disruption model performs equally well compared to other models in terms of event rates.
The appearance of GRB060218 at a redshift $z=0.033$ is consistent with a disruption rate $10^{-3}{\rm yr}^{-1}$ of all stars in a dwarf galaxy.
The rate of low-luminosity GRBs such as our source was estimated to be about $10\%$ of supernova Type Ib/c rate \citep{Toma2007}. A number of shock breakouts
were observed to accompany supernovae at all distances \citep{Woosley:1999la,Mazzali:2006pa,Chevalier:2008fe,Schawinski:2008dw}, and the sample of shock breakouts is likely incomplete.

The long emission duration $2600$~s provides a mass estimate $1\times10^4M_\odot$ for an IMBH within a tidal disruption scenario.
The relativistic shock breakout model also explains the event duration \citep{Nakar:2012aj}.
The extreme smoothness of the spectrum is consistent with either the supernova shock breakout or the jet launched by an IMBH.
The jet launched by a stellar mass central object may produce more time variability.

While the observed absorbed X-ray lightcurve of the event can be fitted with the exponential decay within a shock breakout model \citep{Campana2006},
the source flux corrected for the absorption exhibits $(t-t_0)^{-5/3}$ behavior specific to the tidal disruptions.
Moreover, the full soft X-ray lightcurve including the rise and the decay phases can be fitted well with the accretion rate temporal dependence $\dot{M}(t)$ for a tidal disruption.
The steep flux decay at $\sim6500$~s can be ascribed to the source being more heavily absorbed at late times, while the jet power and the radiation power still scale as $t^{-5/3}$.
The power-law decay is not expected in a shock breakout, while a shallower temporal slope is expected for a jet-powered source \citep{Soderberg:2006na}.
The temporal slope of the afterglow phase, powered by the central engine, is consistent with the accretion rate temporal dependence $t^{-4/3}$ in a RIAF dominated by the disk evolution behavior.

The observed blackbody flux and the blackbody temperature can be readily modeled by the jet photospheric emission
to provide the estimates for the IMBH mass $1\times10^4M_\odot$ and the Lorentz factor $\Gamma\sim3$.
Moreover, these estimates stay consistent between different time slices in our time-resolved spectroscopic analysis.
The rise of the blackbody temperature accompanied by the constant blackbody flux and the dropping total flux leads to the same BH mass and Lorentz factor
over many time slices within the fireball model. The relativistic shock in the wind region of a WR star is able to reproduce the spectrum \citep{Waxman:2007ap}.
However, the properties of the underlying WR star and its wind would be peculiar \citep{Li:2007sh}.

The absence of the jet break is consistent with the wide outflow produced without the pressure support from the surrounding star.
Both the supernova shock breakout and the tidal disruption scenario produce wide outflows.

Strong early optical/UV emission is consistent with coming from the jet front.  The non-relativistic outer shells producing early optical/UV blackbody emission violate
the source energetics \citep{Ghisellini:2007aa} within the supernova shock breakout model.  In addition, the total radiated energy $10^{49.5}$~erg of GRB060218
is somewhat higher than expected for the shock breakouts \citep{Colgate:1974ja} in general.

The host of GRB060218 is a dwarf galaxy, and the dwarf galaxies are expected to have the highest tidal disruption rates.
Only the dwarf galaxies and not the bigger galaxies are expected to have the IMBHs in their nuclei, and thus disrupt the WDs.
We crudely estimate the central BH mass $1\times10^4M_\odot$ based on the stellar mass $10^{7.2}M_\odot$ of the host.
The supernova position is consistent with the center of the host dwarf galaxy, as it should be for the tidal disruption scenario.
Supernovae unrelated to the central BHs generally have large displacements from the galactic centers.
The supernova SN2006aj has the low estimated ejecta mass $M_{\rm ej}=1-2M_\odot$ consistent with the heavy WD.

In sum, the tidal disruption model can explain all features of the source and only the scale of the jet power remains somewhat arbitrary.
Note that the BH mass $1\times10^4M_\odot$ is estimated in three entirely independent ways! A conclusion that GRB060218 was a tidal disruption by the IMBH
in a nucleus of a dwarf galaxy at a distance $150$~Mpc indicates that the IMBHs are abundant in the local Universe.

\section{ACKNOWLEDGEMENTS}\label{sec:acknowledgements}
The authors are thankful to Nathaniel Bulter, John Cannizzo, Coleman Miller, Ramesh Narayan, Eve Ostriker, Enrico Ramirez-Ruiz, and Leslie Sage for useful discussions and comments.
Special thanks to Abderahmen Zoghbi for advice on temporal analysis, Ranjan Vasudevan for general help with XSPEC,
Demosthenes Kazanas and Jonathan McKinney for discussions of the large-scale magnetic field generation.
We thank the anonymous referee for insightful comments, which helped to improve the manuscript.
RVS is supported by NASA Hubble Fellowship grant HST-HF-51298.01, PL is supported by NSF grants 1212433, 1205864, 0941417, 0903973, 0855423, and 0903973,
CSR thanks NASA for support under Astrophysics Theory Program (ATP) grant NNX10AE41G.

\bibliographystyle{apj}

\end{document}